\documentclass[nofootinbib,preprint,aps]{revtex4}
\usepackage[reqno]{amsmath}
\usepackage{amsthm}
\usepackage{amssymb}
\usepackage{array}
\usepackage{longtable}
\usepackage{enumerate}
\usepackage{verbatim}
\usepackage{fancyhdr}
\usepackage{graphicx}
\usepackage{bm}

\usepackage[papersize={8.5in,11in},left=1cm,right=2cm,top=1cm,bottom=1cm,includefoot=true,ignorehead=false,ignoremp=true,driver=dvips,footskip=1cm,headheight=0cm,headsep=0cm]{geometry}
\usepackage[pdftex]{hyperref}
\hypersetup{bookmarksnumbered=true,colorlinks=true,linkcolor=blue,citecolor=magenta}

\providecommand{\C}{{\ensuremath{\mathbb{C}}}}
\providecommand{\al}{\alpha}

\providecommand{\ga}{\gamma}
\providecommand{\Ga}{\Gamma}
\providecommand{\om}{\omega}

\providecommand{\ep}{\epsilon}
\providecommand{\sig}{\sigma}

\DeclareMathOperator{\pRe}{Re} 
\DeclareMathOperator{\pIm}{Im} 
\DeclareMathOperator{\pArg}{Arg} 

\DeclareMathOperator{\Res}{Res}
\newcommand{\EL}[1]{\Biggl\{#1\Biggr\}} 
\newcommand{\GL}[1]{\biggl\{#1\biggr\}} 
\newcommand{\NL}[1]{\bigl\{#1\bigr\}}   

\newcommand{\EC}[1]{\Biggl[#1\Biggr]}   
\newcommand{\GC}[1]{\biggl[#1\biggr]}   
\newcommand{\MC}[1]{\Bigl[#1\Bigr]}     
\newcommand{\NC}[1]{\bigl[#1\bigr]}     

\newcommand{\EP}[1]{\Biggl(#1\Biggr)}   
\newcommand{\GP}[1]{\biggl(#1\biggr)}   
\newcommand{\MP}[1]{\Bigl(#1\Bigr)}     
\newcommand{\NP}[1]{\bigl(#1\bigr)}     

\newcommand{\NB}[1]{\bigl|#1\bigr|}     
\newcommand{\MB}[1]{\Bigl|#1\Bigr|}     
\newcommand{\GB}[1]{\biggl|#1\biggr|}   
\newcommand{\En}[1]{\Biggr|_{#1}}   

\newcommand{\ODo}[1]{\frac{d}{d#1}}

\newcommand{\Do}[2]{\frac{d#1}{d#2}}
\newcommand{\Don}[3]{\frac{d^{#3}#1}{d#2^{#3}}}
\newcommand{\Dp}[2]{\dfrac{\partial{#1}}{\partial{#2}}}
\newcommand{\Dps}[2]{\dfrac{\partial^2{#1}}{\partial{#2}^2}}

\newcommand{\intdef}[4]{\int_{#1}^{#2}{#3\,d#4}}

\newcommand{\intinf}[2]{\int_{-\infty}^{\,\infty}{#1\,d#2}}
\newcommand{\intseminf}[2]{\int_{0}^{\,\infty}{#1\,d#2}}

\newcommand{\ket}[1]{\bigl|#1\bigr\rangle}

\newcommand{\braket}[3]{\bigl\langle#1\bigr|#2\bigl|#3\bigr\rangle}
\newcommand{\Braket}[2]{\bigl\langle#1|#2\bigr\rangle}
\begin{document}
\setcounter{page}{1}
\title
{Quantum decay law: Critical times and the Equivalence of approaches}
\author
{D. F. Ram\'irez Jim\'enez and N. G. Kelkar}
\affiliation{ Departamento de Fisica, Universidad de los Andes,
Cra.1E No.18A-10, Santafe de Bogot\'a, Colombia}
\begin{abstract}
Methods based on the use of Green's functions or the Jost functions and 
the Fock-Krylov method are apparently 
very different approaches to understand the time evolution of unstable states. 
We show that the two former methods are equivalent 
up to some constants and as an outcome find an analytic 
expression for the energy density of states in the Fock-Krylov amplitude in terms 
of the coefficients introduced in the Green's functions and the Jost functions methods. 
This model-independent density is further used to obtain an 
analytical expression for the survival amplitude and study its behaviour at 
large times. Using these expressions, we investigate the origin of the oscillatory 
behaviour of the decay law in the region of the transition from the exponential to 
the non-exponential at large times.
With the objective to understand the failure of nuclear and particle physics 
experiments in observing the non-exponential decay law predicted by quantum mechanics 
for large times, 
we derive analytical formulae for the critical transition time, $t_c$, 
from the exponential to the inverse power law behaviour at large times. 
Evaluating  $\tau_c = \Gamma t_c$ for some particle resonances and 
narrow nuclear states which have been tested experimentally to verify the exponential 
decay law, we conclude that the large time power law in particle and nuclear 
decay is hard to find experimentally. 
\end{abstract}
\maketitle

\section{Introduction}
The decay law of an unstable system  can be shown classically to be of an 
exponential nature but in a quantum mechanical analysis, 
this law is an approximation which fails for short and large times 
\cite{khalfin,urbanowski,fonda,giraldi}. 
The former case is described by a quadratic function in $t$ \cite{shorttime} 
and the latter by an inverse power law in $t$ \cite{fonda}. 
The non-exponential behaviour predicted by quantum mechanics has intrigued 
experimental nuclear and particle physicists who performed experiments 
(see \cite{norman} and references therein) with 
nuclei such as $^{222}$Rn, $^{60}$Co, $^{56}$Mn and measured the decay law 
for several half-lives to disappointingly find only an exponential decay law. 
Even though the non-exponential behaviour at large times was 
confirmed in an experiment measuring the luminescence decays of many dissolved 
organic materials after pulsed laser excitation \cite{rothe}, the failure 
of the nuclear and particle physics experiments raised questions about observation 
such as: 
(i) how long should one wait or in other words, what is the critical transition time 
($\tau_C$) from the exponential to a power law behaviour, 
(ii) does the interaction with the 
environment affect the measurement and (iii) could it be possible that every 
measurement resets the decay to an exponential one, thus making the non-exponential 
behaviour not observable. 
There exist different points of view \cite{fonda,lawrence} and the
above questions still seem to be open for discussions.
Apart from all this, there exist different theoretical formalisms for the quantum 
mechanical treatment of the time evolution and decay of an unstable state 
\cite{Fock1,Sym1,Garcia2,Nakazato,dijknogaPRL,dijknogaPRC,Garcia1,AdvChemGarcia}. In order to at least partly
answer the above questions, it is 
essential to investigate if the different formalisms agree only globally 
on the short and large time behaviour or also in details such as the prediction 
of the critical transition times from the exponential to the power law as well 
as the exponent in the power law behaviour.  
With this objective, in the present work, we investigate some of the most popularly 
known approaches for the calculation of survival probabilities, namely, 
the  method of Garc\'ia-Calder\'on (GC) \cite{Sym1,AdvChemGarcia,AIPGarcia} 
and collaborators \cite{Garcia2,Garcia1,garciaisolated,GarciaPRA76,GarciaPRA88}
which uses the Green's functions, the method of 
W. van Dijk and Y. Nogami (DN) \cite{dijknogaPRL, dijknogaPRC} 
and the Fock-Krylov (FK) method \cite{Fock1} which involves the 
Fourier transform of an energy density.
We show that the seemingly different methods are indeed equivalent and derive 
analytical expressions for the survival amplitudes as well as their large time 
behaviour. A natural continuation of this investigation is to study the 
critical time for the transition from the exponential to the non-exponential 
power law behaviour. Here, we also obtain an analytical expression for the 
critical time and apply it to study the decay of nuclear and particle resonances. 
 
The paper is organized as follows: in Section \ref{SP}, we  
present the basic results of the GC, DN and FK approaches without entering into 
the details of the derivations. In Section \ref{S2}, 
we shall show how the GC and DN approaches lead to the survival amplitude as 
written in the Fock-Krylov method (with a density based on a relation from 
statistical physics) and we shall obtain an 
expression for the energy density of the initial state. 
In Section \ref{S3}, we shall derive an expression for the 
survival amplitude in terms of Incomplete Gamma functions and 
thereby study its behaviour for large $t$. 
Finally, we present the analysis of an isolated resonance and 
obtain an analytic expression to find the critical 
transition time from the exponential to the power law behaviour, in Section \ref{S4.3}. 
Here, we compare our results with an existing work on an isolated resonance  
\cite{garciaisolated}.
Application of the results to realistic resonances in Section \ref{s6}, 
unveils some reasons for the non-observability of non-exponential decay in nuclear 
and particle physics. 
In Section \ref{S4.4}, we present an analysis of the oscillatory transition region 
produced by the interference of the exponential and power law decay at large times. 
We discuss the origin of the oscillatory term and present an expression for the 
modulating function which describes it.
In Section \ref{s7}, we summarize our findings. 

\section{Survival probabilities} 
\label{SP}
In general terms, if ${H}$ is the Hamiltonian of a system and 
its initial state is $\ket{\Psi(0)}$, the state of 
system $\ket{\Psi(t)}$ at a time $t>0$ is given as a solution of the 
Schr\"odinger equation
\begin{equation}\label{E0.1}
i\ODo{t}\ket{\Psi(t)}={H}\ket{\Psi(t)}.
\end{equation}
The quantum decay law is the probability 
(called non-decay or survival probability) that 
the state at time $t$ is in its initial state and is given by, 
$\NB{\Braket{\Psi(0)}{\Psi(t)}}^2=\NB{\braket{\Psi(0)}{e^{-iHt}}{\Psi(0)}}^2$. 
Starting with the survival amplitude $A(t)$, 
one can write it as the projection of the 
state $\ket{\Psi(t)}$ on the state  $\ket{\Psi(0)}$:
\begin{equation}\label{E0.3}
A(t)=\Braket{\Psi(0)}{\Psi(t)}=\braket{\Psi(0)}{e^{-iHt}}{\Psi(0)},
\end{equation}
and the survival probability mentioned above is simply
\begin{equation}\label{E0.4}
P(t)=|A(t)|^2.
\end{equation}
Both the survival amplitude and the survival probability are equal to one when $t=0$ because $A(0)=\Braket{\Psi(0)}{\Psi(0)}=1$ and from \eqref{E0.4}, $P(0)=|A(0)|^2=1$.

We note here that in the context of the time evolution of an unstable state, 
a widely discussed quantity is also the so called ``nonescape probability" which is 
essentially the probability that the particle remains confined inside the interaction 
or potential region after a time $t$. Though the two concepts of survival and nonescape 
probabilities are closely related, there are instances when they could be significantly 
different \cite{dijkPRE93}.  We refer the reader to 
\cite{dijkPRE93,Garcia2,CavalPRL80,garciaPRL80,dijkPRL901,garciaPRL90,GarciaPRA76} 
for some interesting discussions on this topic.
\subsection{Fock-Krylov method}
A popular method for computing the survival amplitude is the 
Fock-Krylov (FK) approach \cite{Fock1}. 
In general, this method involves expanding the initial 
state in eigenstates of a complete set of observables which commute with the 
Hamiltonian. If we let $\NL{{H},{\beta}}$ to be this set of 
observables and let $\ket{E,b}$ be an eigenstate of them:
\begin{align}
{H}\ket{E,b}&=E\ket{E,b}\label{E0.5},\\
{\beta}\ket{E,b}&=b\ket{E,b}\label{E0.6}\, , 
\end{align}
then the normalized initial state can be expanded in this basis as
\begin{equation}\label{E0.7}
\vert \Psi(0) \rangle =\int_{E_{\text{min}}}{dE\int{db\,\ket{E,b}\Braket{E,b}{\Psi(0)}}}.
\end{equation}

Let us now consider an intermediate unstable state (resonance) formed in
a scattering process such as $A + a \rightarrow R^* \rightarrow A + a$.
Since the initial unstable state, $\vert \Psi (0) \rangle$, cannot be an
eigenstate of the (hermitian) Hamiltonian, an expansion 
as in (\ref{E0.7}) 
(assuming a continuous spectrum) in terms of the 
energy eigenstates of the decay products $A$ and $a$ can be considered to express 
$\vert \Psi(0) \rangle$ as 
\begin{equation}\label{psi}
\vert \Psi (0) \rangle =  \int dE \,\,a(E)\, \vert E \rangle
\end{equation}
where $\vert E \rangle$ is the eigenstate and $E$ the total energy of the
system $A + a$. 
Substituting now for $\vert \Psi(0) \rangle$ in (\ref{E0.3}), we get,
\begin{align}
A(t) &= \int dE'\, dE \,\,a^*(E')\, a(E)\, \langle E' 
\vert e^{(-iHt)} \vert E \rangle \notag\\
&= \int dE' \,dE \,\,a^*(E') \,a(E)\, e^{(-iEt)} \,\delta(E - E')\\
&= \int \,dE\, |a(E)|^2\,e^{-iEt}
\end{align}
The proper normalization of $\Psi$ tells us that $|a(E)|^2$ should
have the dimension of ($1/E$) and hence can be associated with an energy density of
states. Thus, in the Fock-Krylov method,
\begin{equation} \label{fock}
A(t)=\int_{E_{\rm th.}}^{\infty} dE \,\,
\rho(E) \,e^{-iEt}
\end{equation}
where $E_{\rm th.}$ is the minimum sum of the masses of the decay products.

\subsection{Statistical physics based approach}\label{phenomenological}  
The advantage of the FK method is that it is not necessary to solve 
the Schr\"odinger equation and in cases where one does not know $\ket{\Psi(0)}$, 
one can proceed to evaluate $A(t)$ if the ``spectral function" or the energy 
distribution of the resonant state is known. Such an approach was given in 
\cite{Kelkar1,Kelkar2} in order to analyze realistic cases of nuclear and particle 
resonances. The authors noted that theoretically, many different forms of 
$\rho(E)$ are available but they may not necessarily have a connection with experiments. 
One of the experimental signatures for the existence of a resonance is the sharp 
jump in the phase shift $\delta(E)$, as a function of energy. 
The energy derivative of the 
phase shift displays the typical Lorentzian form associated with a resonance 
\cite{ourarxiv} and has different interpretations. One of its first appearances in 
relation with resonances was in the definition of Wigner's time delay \cite{wigner55}. 
Though Wigner's work dealt with the single channel case, the energy derivative, 
$d\delta_l(E)/dE$, for a resonance occurring in the $l^{th}$ partial wave in 
scattering can be shown to be {\it 
the difference between the time spent by the interacting particles with and 
without interaction in a given region of space} \cite{smith,meandMPRA,mePRL}.
This interpretation led the authors in \cite{Kelkar1,Kelkar2} to find the connection 
between $d\delta_l(E)/dE$ and $\rho_l(E)$ for the density of states in a resonance. 

In calculating the second virial coefficients $B$ and $C$ for the equation of states 
in a gas, $pV=RT[1 +B/V + C/V^2 +...]$, Beth and Uhlenbeck
\cite{bethuhl} (the derivation of their result is reproduced
in \cite{huang}, see also \cite{dashen1,dashen2})
found that the difference between the density of states with interaction, 
$n_l$, and without, $ n_l^{(0)}$, is given by the derivative of the scattering 
phase shift $\delta_l$ as,
\begin{equation} \label{densitystates}
\rho_l^{BU}(E) = n_l(k) - n_l^{(0)}(k)=\frac{2l+1}{\pi}\,\frac{d\delta_l(E)
}{dE}
\end{equation}
where $k$ and $E$ are the momentum and energy in the
centre-of-mass system of the scattering particles,
respectively. If a resonance is formed during the scattering process,
$E$ becomes the energy of the resonance in its rest frame.
In the absence of interaction, since no resonance can be produced,
one would expect the density of states $\rho(E)$ to be zero. 
If the interaction is switched off, $n_l$ will tend to $n_l^{(0)}$ from above.
Therefore, the authors concluded that, 
as long as $n_l-n_l^{(0)} \ge 0$ (which is at least the case for an
isolated resonance), one can write for the continuum probability density of
states of the decay products in a resonance,
\begin{equation}\label{rhoanddelta}
\rho^{BU}_l(E) = {\rm const.} \,\frac{d \delta_l (E)}{d E}\, . 
\end{equation}
Finally, using the phase shift values extracted from scattering experiments, 
the authors calculated the survival amplitude, Eq. (\ref{fock})
with the substitution of Eq. (\ref{rhoanddelta}) as 
\begin{equation} \label{phenoA}
A_l(t)=\int_{E_{\rm th.}}^{\infty} dE \,\frac{d \delta_l (E)}{d E} 
\,e^{-iEt}\, . 
\end{equation}
Analytical expressions for the above have been provided in a recent work 
\cite{ourarxiv} with the use of the Mittag-Leffler theorem. 

Before we proceed to the next subsections, 
let us clarify the notation used in this work. For mathematical simplicity,
we set, $2m=\hbar^2 = 1$ and hence $k^2 = E$. 
For a resonance pole given by $E_{r} - i \Gamma_r/2$ 
in the complex energy plane, $k^2_{r} = \epsilon_{r} - i\Gamma_{r}/2$ where
$\epsilon_{r} = E_{r} - E_{th}$, with $E_{th}$ being the threshold energy
(or the sum of the masses of the decay products of the resonance).
Having shifted the energies by an amount $E_{th}$, the lower limit on the 
integral for the survival amplitude  
will be 0 instead of the threshold energy $E_{th}$.   

\subsection{Green's function method}
\label{S1}
Another method for obtaining the survival amplitude is 
to solve \eqref{E0.1} using Green's functions. 
Finding the Green's function may be a laborious undertaking, however,  
there exists 
an elegant approach proposed by Garcia-Calderon (GC) \cite{Sym1} (and followed up 
in \cite{Garcia1,Garcia2,Sym2,AdvChemGarcia,GarciaPRA88}) 
which overcomes this difficulty.   
The GC approach uses resonant states for calculating the Green's function, 
the corresponding wave function and the survival amplitude. 
Since those resonant states are intimately connected 
with the poles of the $S$-matrix (see \cite{ourarxiv} and references therein for 
realistic studies and 
\cite{kangoller} for different 
pole structures in scattering), 
it is possible to express $A(t)$ analytically, in particular, 
in terms of error functions. In what follows, we shall briefly discuss the 
GC method and recommend Refs \cite{Garcia1,Garcia2,Garcia3,AdvChemGarcia,AIPGarcia} 
to the interested reader for details of the formalism. 

The GC method is based on building the wave function through the 
Green's function of a system using resonant states. 
Let us consider the system to be a particle of mass $m$ without spin  
moving under the influence of a central potential $V(r)$ of finite range $R$. 
At time $t$ = 0, this system is described by an 
initial wave function $\psi(r,0)$ which is zero in $r>R$, i. e., the particle is confined in the region $r<R$. 
If $\psi(r,t)$ is the state of the system after a time $t$ (here $\psi(r,t)$ 
is actually the wave function, $\Psi$, times $r$), 
for S-waves, it must satisfy 
the Schr\"odinger equation:
\begin{equation}\label{E1.1}
-\Dps{\psi(r,t)}{r}+V(r)\psi(r,t)=i\Dp{\psi(r,t)}{t}.
\end{equation}
Using Green's functions, 
it is possible to show that the wave function $\psi(r,t)$ can be written as
\begin{equation}\label{E1.2}
\psi(r,t)=\sum_n{C_n(k_n)u_n(r,k_n)M(k_n,t)},
\end{equation}
where the sum is over all poles of the $S$-matrix. 
The authors in \cite{Garcia1,AdvChemGarcia} make use of the fact that for a 
finite range interaction, the outgoing Green's function as a function of 
the momentum, $k$, can be extended analytically to the whole complex $k$ 
plane where it has an infinite number of poles. 
As is well known, purely imaginary poles in 
the upper half of the complex $k$ plane correspond to bound states and 
those in the lower half plane correspond to virtual states. 
Complex poles are however found only in the lower half of the complex 
$k$ plane and corresponding to every pole $k_n = a_n - ib_n$ 
($a_n$, $b_n > 0$), there exists due to time reversal invariance, a 
complex pole, $k_{-n}$, situated symmetrically with respect to the 
imaginary axis, i.e., $k_{-n} = - k_n^*$. In \cite{Garcia1,AdvChemGarcia}, the authors 
considered examples with potentials having no bound states so that 
all poles were located only in the lower half of the complex $k$ plane. 
 
Coming back to \eqref{E1.2}, $M(k_n,t)$ is the integral
\begin{equation}\label{E1.3}
M(k_n,t)=\frac{i}{2\pi}\intinf{\frac{e^{-itx^2}}{x-k_n}}{x},
\end{equation}
$u_n(r,k_n)$ is the resonant state associated with the pole $k=k_n$ and 
is the solution of the differential equation \cite{Garcia3}
\begin{equation}\label{E1.4}
\Don{u_n(r,k_n)}{r}{2}+\NC{V(r)-k_n^2}u_n(r,k_n)=0,
\end{equation}
with boundary conditions
\begin{align}
u_n(0,k_n)&=0,\label{E1.5}\\
\Do{u_n(r,k_n)}{r}\En{r=R}&=ik_nu_n(R,k_n)\label{E1.6},
\end{align}
and satisfies the normalization condition 
\footnote{This normalization is equivalent to the one originally proposed by 
Zel'dovich \cite{zeldovich}. A similar form was also found in 
\cite{Gareev}. We refer the reader to \cite{Garcia3} for more references.}
\begin{equation}
\intdef{0}{R}{u^2(r,k_n)}{r}+\frac{i}{2k_n}u_n^2(R,k_n)=1.
\end{equation}
which as given in \cite{Garcia3} follows from the residue at a
complex pole $k_n$ of the outgoing Green's function to the problem. 
The coefficient $C_n(k_n)$ is given by
\begin{equation}\label{E1.7}
C_n(k_n)=\intdef{0}{R}{\psi(r,0)u_n(r,k_n)}{r}.
\end{equation}
The survival amplitude, in this case is given by, 
\begin{equation}\label{E1.8}
A(t)=\Braket{\Psi(r,0)}{\Psi(r,t)}=\int_{0}^{R}{\psi^*(r,0)\psi(r,t)\,dr}=\sum_n{C_n(k_n)\bar{C}_n(k_n)M(k_n,t)},
\end{equation}
where the coefficient $\bar{C}_n(k_n)$ is:
\begin{equation}\label{E1.9}
\bar{C}_n(k_n)=\intdef{0}{R}{\psi^*(r,0)u_n(r,k_n)}{r}.
\end{equation}
Each pair of coefficients $C_n(k_n)$ and $\bar{C}_n(k_n)$ for a given 
$n$ satisfy certain properties (see \cite{Garcia1} for details). 
In the appendix \ref{A}, we use the steepest descent method for showing that, 
for large $t$ the survival amplitude given by \eqref{E1.8} is equal to:
\begin{equation}\label{E1.10}
A(t)=-\frac{1}{\sqrt{4\pi}}e^{\,i\pi/4}\pIm{\GC{\sum_p{\frac{C_p(k_p)\bar{C}_p(k_p)}{k_p^3}}}}t^{-3/2}+O(t^{-5/2}),
\end{equation}
where $\NL{k_p}$, $p=1,2,\dotsc$ are the fourth-quadrant poles of the S-matrix in the complex $k$ plane.
For large $t$, the survival amplitude is proportional to $t^{-3/2}$ and the survival probability is proportional to $t^{-3}$. This result for $l$ = 0 is consistent with 
the expectation of $t^{2l+3}$ (for the $l^{th}$ partial wave) 
in literature \cite{bogda,fonda,Kelkar1,Kelkar2}. It is also consistent with the 
density given by $\rho_l(E) \propto {d \delta_l (E)/d E}$ since one expects the 
phase shift to behave as $\delta_l \sim k^{2l +1}$ near threshold which eventually leads 
to the above power law at large times (see Sections 4.3 and 5.2 in \cite{ourarxiv}). 
In the experimental observation of the non-exponential decay \cite{rothe}, however, 
the exponent was found to vary between -2 to -4. 
The experimental observation was made with complex organic systems which are 
not spherically  symmetric and hence it is not surprising  
that other powers of time are exhibited. 
We also note that in \cite{onleykumar}, 
within a model of a two level system coupled to the 
continuum, the authors found an exponent of -4. 

A small note regarding the steepest descents method is in order here before closing this 
subsection. 
This method has been used earlier in \cite{Sym1,GarciaPRA76,
AdvChemGarcia} in the 
context of arriving at the above result but in a somewhat different manner as compared 
to the present work where it is used to directly evaluate $A(t)$.
The authors in \cite{GarciaPRA76} for example, use this method in order to obtain the 
retarded Green's function, $g(r,r^{\prime};t)$, entering into the definition of the 
time evolved wave function, namely, $\psi(r,t) = \int_0^R g(r,r^{\prime};t) 
\psi(r^{\prime},0) dr^{\prime}$, which eventually defines the survival amplitude.
The contours of integration in \cite{GarciaPRA76} and in the present work are hence 
also different. 

\subsection{Jost and Moshinsky functions method}\label{sectiondijk}
In an attempt to obtain the expression for the wave function of a decaying quantum 
system, the authors W. van Dijk and Y. Nogami (DN) in Ref. \cite{dijknogaPRL}, 
proposed a method which involved the description of the wave function as a linear 
combination of the the Moshinsky functions, $M(k,r,t)$ \cite{moshPRA8488}, each of 
which is associated with a pole of the scattering matrix, {\bf S}. 
In a follow-up work \cite{dijknogaPRC}, 
the authors used this formalism to study the survival and 
nonescape probabilities of decaying quantum systems. In this subsection, we shall 
describe the DN approach for the evaluation of survival probabilities in some 
detail, in order to later compare it with the GC and FK approaches discussed before. 

The authors in \cite{dijknogaPRC} begin by considering 
the case of $S$-wave unstable states and attempt to find a solution of the time 
dependent Schr\"odinger equation with a  central potential V(r) of finite range 
$R$ and an initially normalized wave function $\psi(r,0)$ confined to $r<R$, 
i.e., $\psi(r,0)=0$ for $r>R$. 
The scattering solutions are expressed in terms of Jost functions such that for the 
case of no bound states,
\begin{equation}\label{E1}
 \psi(r,t)=\frac{2}{\pi}\intseminf{\frac{k^2}{|f(k)|^2}c(k)u(k,r)e^{-ik^2t}}{k},
 \end{equation}
where $k > 0$ and $k^2$ is the corresponding energy. 
Here, $\psi(r,t)$ is the wave function, $\Psi$, times $r$. $c(k)$ is given as, 
\begin{equation}\label{E2}
c(k)=\intdef{0}{R}{\psi(r,0)u(k,r)}{r},
\end{equation}
with $u(k,r)$ being the function defined as
\begin{equation}\label{E3}
u(k,r)=\frac{1}{2ik}\NC{f(k)f(-k,r)-f(-k)f(k,r)}\,. 
\end{equation}
$f(k,r)$ is the Jost solution of the time independent Schr\"odinger 
equation \cite{joachain} (with potential V(r)) and 
$f(k)$ is the Jost function related to it as $f(\pm k) = f(\pm k,0)$. 
The function $u(k,r)$ is real and $u(k,r)$ and $c(k)$ are both entire and even 
in the parameter $k$. This function is normalized such that
\begin{equation}
\intseminf{u(k,r)u(k',r)}{r}=\frac{\pi}{2k^2}|f(k)|^2\delta(k-k').
\end{equation}
Using \eqref{E1} the survival amplitude is written as, 
\begin{equation}\label{E5}
A(t)=\intseminf{\psi^*(r,0)\psi(r,t)}{r}=\frac{2}{\pi}\int_{0}^{\infty}{dk\,\frac{k^2}{|f(k)|^2}c(k)e^{-ik^2t}\GC{\intdef{0}{R}{\psi^*(r,0)u(k,r)}{r}}}\, .
\end{equation}

Before proceeding to the comparison of approaches reviewed in this section, 
we recall an older work \cite{cavalcanti}  
on the complex energy eigenfunctions 
(as those given by Eqs (\ref{E1.4}) - (\ref{E1.6})) 
where the effectiveness of this method, in spite of 
several shortcomings has been discussed. 
Another work worth 
mentioning in the context of the present investigations is Ref. \cite{GarciaPhysScripta} 
where the authors performed a comparison of the Hermitian and non-Hermitian formulation 
for the time evolution of quantum decay and showed that they lead to an identical 
description for a large class of well-behaved potentials.

\section{Energy density of the initial state}\label{S2}
Having introduced the different approaches for the calculation of survival 
amplitudes, we shall now examine the expressions, Eq. (\ref{phenoA}), 
(\ref{E1.8}) and (\ref{E5}) to obtain a definition of the energy density of states in the GC and 
DN formalisms and compare the survival probabilities in these two approaches with 
that of the frequently used Fock-Krylov method. 

\subsection{GC formalism}\label{rhoEGC}
We begin by writing the integral $M(k_n,t)$ as
\begin{align}
M(k_n,t)
&=\frac{i}{2\pi}\intseminf{\frac{e^{-itx^2}}{x-k_n}}{x}
-\frac{i}{2\pi}\intseminf{\frac{e^{-itx^2}}{x+k_n}}{x}\notag\\
&=\frac{i}{2\pi}\intseminf{\GP{\frac{1}{x-k_n}-\frac{1}{x+k_n}}e^{-itx^2}}{x}\notag\\
&=\frac{1}{2\pi i}\intseminf{\frac{2k_n}{k_n^2-x^2}e^{-itx^2}}{x},\label{E2.1}
\end{align}
and making the change of variable $E=x^2$, we have 
\begin{equation}\label{E2.2}
M(k_n,t)=\frac{1}{2\pi i}\intseminf{\frac{k_n}{\sqrt{E}\NP{k_n^2-E}}e^{-itE}}{E}.
\end{equation}
Since
\[
\frac{k_n}{k_n^2-E}=\frac{1}{k_n}\GP{1+\frac{E}{k_n^2-E}},
\]
the integral takes the form:
\begin{equation}\label{E2.3}
M(k_n,t)=\frac{1}{2\pi i}\intseminf{\frac{\sqrt{E}}{k_n\NP{k_n^2-E}}e^{-itE}}{E}
+\frac{1}{2\pi ik_n}\intseminf{\frac{e^{-itE}}{\sqrt{E}}}{E}.
\end{equation}
Substituting \eqref{E2.3} in \eqref{E1.8}, we obtain:
\begin{multline}\label{E2.4}
A(t)=\intseminf{\EC{\frac{1}{2\pi i}\sum_n{C_n(k_n)\bar{C}_n(k_n)\frac{\sqrt{E}}{k_n\NP{k_n^2-E}}}}e^{-itE}}{E}\\
+\frac{1}{2\pi i}\GC{\sum_n{\frac{C_n(k_n)\bar{C}_n(k_n)}{k_n}}}\intseminf{\frac{e^{-itE}}{\sqrt{E}}}{E}.
\end{multline}
The last term is zero because of the properties of the coefficients $C_n$. 
The final form of the survival amplitude is: 
\begin{equation}\label{E2.5}
A(t)=\intseminf{\EC{\frac{1}{2\pi i}\sum_n{C_n(k_n)\bar{C}_n(k_n)\frac{\sqrt{E}}{k_n\NP{k_n^2-E}}}}e^{-itE}}{E}.
\end{equation}
We can see that $A(t)$ is the Fourier transform of the  
series given in the square brackets. In other words, the GC approach 
leads to a survival amplitude which is very similar in form to that of 
the Fock-Krylov method. Comparing Eq. \eqref{E2.5} 
with the FK amplitude given in Eq. \eqref{fock}, we consider identifying 
the quantity in square brackets with the energy density $\rho(E)$ 
of the initial state and write 
\begin{equation}\label{E2.6}
\rho^{GC}(E)={\frac{1}{2\pi i}\sum_n{C_n(k_n)\bar{C}_n(k_n)\frac{\sqrt{E}}{k_n\NP{k_n^2-E}}}},
\end{equation}
If we perform the last sum with poles of the fourth quadrant only, this energy 
density can be written as
\begin{equation}\label{E2.7}
\rho^{GC}(E)={\frac{1}{\pi}\pIm{\sum_p{C_p(k_p)\bar{C}_p(k_p)\frac{\sqrt{E}}
{k_p\NP{k_p^2-E}}}}}.
\end{equation}

\subsection{DN formalism}\label{rhoEDN}
Let us start by considering the integral in the square brackets 
in (\ref{E5}). It is the complex conjugate of $c(k)$ given by \eqref{E2}. Thus
 \begin{equation}\label{E6}
A(t)=\frac{2}{\pi}\intseminf{k^2\frac{|c(k)|^2}{|f(k)|^2}e^{-ik^2t}}{k}.
\end{equation}
Performing a change of variable $k^2=E$, we have:
\begin{equation}\label{E7}
A(t)=\intseminf{\frac{\sqrt{E}}{\pi}\GB{\frac{c(\sqrt{E})}{f(\sqrt{E})}}^2\,e^{-iEt}}{E}.
\end{equation}
Comparing the above expression with the Fock-Krylov amplitude, the energy density 
in the DN formalism is given by, 
\begin{equation}\label{E8}
\rho^{DN}(E)=\frac{\sqrt{E}}{\pi}\GB{\frac{c(\sqrt{E})}{f(\sqrt{E})}}^2.
\end{equation}

If we consider the integrand in (\ref{E6}) without the exponential part 
$e^{-ik^2t}$, then using \eqref{E2} and the property of the Jost 
function $f^*(k)=f(-k)$ for real $k$, we get:
\begin{equation}\label{E9}
\varrho(k)\equiv\frac{2}{\pi}\,k^2\,\frac{|c(k)|^2}{|f(k)|^2}=
\frac{2}{\pi}\,\intdef{0}{R}{\intdef{0}{R}{\psi(r,0)\psi^*(r',0)\GC{k^2\frac{u(k,r)u(k,r')}{f(k)f(-k)}}}{r'}}{r}.
\end{equation}
Now, taking the function in square brackets (let us call it $I(k,r,r')$) and  
considering the definition of the S-matrix in terms of the Jost functions, 
\begin{equation}\label{E11}
S(k)=\frac{f(k)}{f(-k)},
\end{equation}
we get, 
\begin{equation}\label{E12}
I(k,r,r')=-\frac{1}{4}\NC{S(k)f(-k,r)-f(k,r)}\NC{f(-k,r')-f(k,r')/S(k)}.
\end{equation}
Taking into account that $\NL{k_p}$, $p=1,2,\dotsc$ are the poles of S-matrix in 
the fourth-quadrant of the complex $k$ plane, 
the S matrix has additional poles $\NL{-k_p^*}$ and 
zeros $\NL{-k_p}$ and $\NL{k_p^*}$, where all these zeros and poles are simple \cite{Zeldovich2}.
Thus, the poles of the function $I(k,r,r')$ correspond to the
poles and zeros of the S-matrix.
If $b_p$ are the residues of the S-matrix in the fourth-quadrant, the residues
corresponding to its poles of the third-quadrant are $-b_p^*$ \cite{ourarxiv} 
while the residues of the inverse of the S-matrix, 
in terms of $b_p$ are (see Appendix \ref{C}):
\begin{align}
\Res\NC{1/S(k),k=-k_p}&=-b_p,\label{E13}\\
\Res\NC{1/S(k),k=k_p^*}&=b_p^* \,.\label{E14}
\end{align}
Thus, the residues of the function $I(k,r,r')$ can be expressed in terms of the 
residues of the S-matrix. If we call $\iota(k_p,r,'r)$ the residues of this function 
corresponding to the poles of the fourth-quadrant, then:
\begin{align}
\Res\NC{I(k,r,r'),k=k_p}&=-\frac{1}{4}b_p{f(-k_p,r)f(-k_p,r')}\equiv \iota(k_p,r,r'),\label{E15}\\
\Res\NC{I(k,r,r'),k=-k_p^*}&=\frac{1}{4}b_p^*{f^*(-k_p,r)f^*(-k_p,r')}=-\iota^*(k_p,r,r'),\label{E16}\\
\Res\NC{I(k,r,r'),k=-k_p}&=\frac{1}{4}b_p{f(-k_p,r)f(-k_p,r')}=-\iota(k_p,r,r'),\label{E17}\\
\Res\NC{I(k,r,r'),k=k_p^*}&=-\frac{1}{4}b_p^*{f^*(-k_p,r)f^*(-k_p,r')}=\iota^*(k_p,r,r').\label{E18}
\end{align}
In the calculation of these residues, we used the property $f^*(-k^*,r)=f(k,r)$ 
for complex $k$ \cite{Sitenko} (see also \cite{rakitelander} for a 
discussion on the use of Jost functions in bound and resonant state problems). 
From the Mittag-Leffler theorem
\footnote{If the only singularities of a meromorphic function $f(z)$ are the 
simple poles $z=a_1,a_2,\dotsc$ such that $|a_1| \le |a_2| \le ...$,  
with residues $b_1,b_2\dotsc$ respectively, and if 
$C_N$ is a circumference of radius $R_N$ which contains N poles of the function $f(z)$ 
(and does not pass through any of the remaining poles), i.e., $|a_N|<R_N<|a_{N+1}|$, 
and on $C_N$, $|f(z)|<M$, where $M$ is not dependent on $N$, then
\[
f(z)=f(0)+\lim_{N\to\infty}{\sum_{n=1}^{N}{b_{n}\GL{\frac{1}{z-a_n}+\frac{1}{a_n}}}}+
\lim_{N\to\infty}{\frac{z}{2\pi i}\oint_{C_N}{\frac{f(\zeta)}{\zeta(\zeta-z)}\,d\zeta}}
=f(0)+\sum_{n=1}^{\infty}{{\frac{b_nz}{a_n(z-a_n)}}}.
\]
This theorem is known as the Mittag-Leffler theorem \cite{MittagL}.}
and taking into 
account that $I(0,r,r')=0$, we have:
\begin{equation}\label{E19}
I(k,r,r')=4k^2\,\pRe{\EC{\sum_p{\frac{\iota(k_p,r,r')}{k_p(k^2-k_p^2)}}}}.
\end{equation}
Finally, after some lengthy algebra 
(see Appendix \ref{D}), it is possible to write $\varrho(k)$ in the 
following form:
\begin{equation}\label{E24}
\varrho(k)=
\frac{2}{\pi}\,k^2\,\pRe{\sum_p{\frac{ia_p(k_p)}{k_p(k_p^2-k^2)}}},
\end{equation}
where the coefficients $a_p(k_p)$ are given by, 
\begin{equation}\label{E23}
a_p(k_p)\equiv 4i\intdef{0}{R}{\intdef{0}{R}{\psi(r,0)\psi^*(r',0)\iota(k_p,r,r')}{r'}}{r}.
\end{equation}
Noting the definition of $\varrho(k)$ in (\ref{E9}) and 
substituting \eqref{E24} in \eqref{E6}, the survival amplitude in the DN formalism 
becomes, 
\begin{equation}\label{E25}
A(t)=\intseminf{\frac{2}{\pi}\pRe{\GC{k^2\sum_p{\frac{ia_p(k_p)}{k_p(k_p^2-k^2)}}}}e^{-ik^2t}}{k} \, , 
\end{equation}
which, after a change of variable $E=k^2$, can be expressed as,
\begin{equation}\label{E26}
A(t)=\intseminf{\frac{1}{\pi}\pRe{\GC{\sqrt{E}\sum_p{\frac{ia_p(k_p)}{k_p(k_p^2-E)}}}}e^{-iEt}}{E}=\intseminf{\frac{1}{\pi}\pIm{\GC{\sum_p{\frac{a_p(k_p)}{k_p}\frac{\sqrt{E}}{k_p^2-E}}}}e^{-iEt}}{E}\, , 
\end{equation}
so that 
\begin{equation}\label{E26p}
\rho^{DN}(E) = \frac{1}{\pi}\pIm{\GC{\sum_p{\frac{a_p(k_p)}{k_p}\frac{\sqrt{E}}{k_p^2-E}}}}. 
\end{equation}
The above expression for the survival amplitude is the same as that in Eq. (\ref{E2.7}) 
up to the constants $C_p(k_p) \bar{C}_p(k_p)$ 
and $a_p(k_p)$. Note however that
there is a subtle difference between the constants of the GC and DN formalism.
$C_p(k_p)\bar{C}_p(k_p)$ of the GC formalism depend solely on the 
resonant poles $k_p$. 
However, the constants $a_p(k_p)$ which 
apparently depend only on $k_p$, in principle depend on all other existing 
poles through their dependence on the residues $\iota(k_p,r,r')$ 
(see Eqs (\ref{E15}) and (\ref{EC37p})). Though in practice such a calculation may not 
be feasible, under certain conditions, it is possible to use an approximate 
solution as given in Appendix \ref{C}. 

\subsection{Comparison of the GC and DN coefficients}
The coefficients in the GC formalism written as a double integral:
\begin{equation}\label{E27}
C_p(k_p)\bar{C}_p(k_p)=\int_{0}^{R}{\int_{0}^{R}{\psi(r,0)\psi^*(r',0)u_p(r,k_p)u_p(r',k_p)}\,drdr'},
\end{equation}
where $R$ is the range of the potential, are not equal to the coefficients $a_p(k_p)$ unless
\begin{equation}\label{E28}
u_p(r,k_p)=\sqrt{\frac{b_p}{i}}f(-k_p,r) \, .
\end{equation}
The above expression is deduced by substituting the definition of $\iota(k_n,r,r')$ in the 
integral \eqref{E23} and comparing with \eqref{E27}. 
In principle, this result shows that the resonant state associated with the fourth-quadrant pole 
$k=k_p$ which is also a pole of the S-matrix may be computed in terms of the residues of 
the S-matrix at the corresponding pole and the Jost function.

From the Riemann-Lebesgue theorem we know that $\varrho(k)\to0$ when $k\to\infty$. This implies that
\begin{equation}\label{E29}
\lim_{k\to\infty}{\varrho(k)}=\lim_{k\to\infty}{\frac{2}{\pi}\,k^2\,\pRe{\sum_p{\frac{ia_p(k_p)}{k_p(k_p^2-k^2)}}}}=0\quad\Rightarrow\quad\pIm{\sum_p{\frac{a_p(k_p)}{k_p}}}=0.
\end{equation}
Since $A(0)=1$, from \eqref{E26} we have that:
\begin{equation}\label{E30}
\intseminf{\frac{1}{\pi}\pIm{\GC{\sum_p{\frac{a_p(k_p)}{k_p}\frac{\sqrt{E}}{k_p^2-E}}}}}{E}=1.
\end{equation}
However, 
\begin{equation}\label{E31}
\frac{\sqrt{E}}{k_p^2-E}=\frac{1}{\sqrt{E}}\GP{\frac{k_p^2}{k_p^2-E}-1}.
\end{equation}
Using the condition \eqref{E29}, Eq. \eqref{E30} takes the form:
\begin{equation}\label{E32}
\frac{1}{\pi}\pIm{\sum_p{\frac{a_p(k_p)}{k_p}\intseminf{\frac{1}{\sqrt{E}}\GP{\frac{k_p^2}{k_p^2-E}-1}}{E}}}=\frac{1}{\pi}\pIm{\sum_p{{k_p}{a_p(k_p)}\int_{0}^{\infty}{\frac{dE}{\sqrt{E}\NP{k_p^2-E}}}}}=1.
\end{equation}
Since the integral in the last equation is equal to $i\pi/k_p$, \eqref{E32} 
reduces to \footnote{The integral was calculated following this theorem: If $f(z)$ is a single-valued analytic function in the domain $0<\pArg{z}<2\pi$, except for a finite number of singularities $z_k$, $k=1,\dotsc n$ not lying on the positive real axis and let $z=\infty$ be a zero of order not lower than first of the function $f(z)$, then
\[
\intseminf{x^{\al-1}f(x)}{x}=\frac{2\pi i}{1-e^{\,2\pi i\al}}\sum_{k=1}^{n}
{\Res{\NC{z^{\al-1}f(z),z=z_k}}},
\]
where $0<\al<1$ \cite{Tijonov}.}
\begin{equation}\label{E33}
\pIm{i\sum_p{a_p(k_p)}}=\pRe{\sum_p{a_p(k_p)}}=1.
\end{equation}
The properties \eqref{E29} and \eqref{E33} satisfied by the coefficients 
$a_p(k_p)$ are the same as those satisfied by $C_p(k_p)\bar{C}_p(k_p)$. 

Since the energy density 
and hence the survival amplitude in the GC and DN formalisms have been 
shown in the previous section to be equivalent up to the constants appearing in 
Eqs (\ref{E2.5}) and (\ref{E26}), it is convenient to write both equations in one 
compact expression before computing the survival amplitude and other 
quantities of interest. Thus, if we define the coefficient $\gamma_p(k_p)$ as:
\begin{equation}\label{coeff}
\ga_p(k_p)=
\begin{cases}
C_p(k_p)\bar{C}_p(k_p), & \text{for GC formalism},\\
a_p(k_p), & \text{for DN formalism},
\end{cases}
\end{equation}
then both the energy densities can be written in a common form as:
\begin{equation}\label{density}
\rho(E) = \frac{1}{\pi}\pIm{\GC{\sum_p{\frac{\ga_p(k_p)}{k_p}\frac{\sqrt{E}}{k_p^2-E}}}}. 
\end{equation} 

The coefficients $\ga_p(k_p)$ satisfy the same properties as $C_p(k_p)\bar{C}_p(k_p)$ and $a_p(k_p)$, i.e., 
\begin{align}
\pRe{\sum_p{\ga_p(k_p)}}&=1,\label{coef1}\\
\pIm{\sum_p{\frac{\ga_p(k_p)}{k_p}}}&=0.\label{coef2}
\end{align}

\subsection{Energy density of an isolated resonance}
In the case of an isolated resonance, with a 
pole at say $k_r$, the conditions on the coefficients $\gamma_r$ given by 
Eqs. \eqref{coef1} and \eqref{coef2} are reduced to
\begin{align}
\pRe{\ga_r(k_r)}&=1,\label{E3.23}\\
\pIm{\frac{\ga_r(k_r)}{k_r}}&=0\label{E3.24}.
\end{align}
leading to 
\begin{equation}\label{E3.26}
\ga_r(k_r)=1+i\frac{\pIm{\NP{k_r}}}{\pRe{\NP{k_r}}}=\frac{k_r}{\pRe{\NP{k_r}}}.
\end{equation}
This equation was first deduced in \cite{garciaisolated} for the GC formalism and is 
also valid for the DN formalism. Replacing the above in \eqref{E2.7}, we obtain the 
energy density for an isolated resonance, 
\begin{equation}\label{E2.7p}
\rho^{GC}_{iso}(E)={\frac{1}{\pi}\pIm{\frac{\sqrt{E}}
{k_r\NP{k_r^2-E}}}} \, + \, {\frac{1}{\pi} \pIm \,\biggl [ 
{ i\frac{\pIm{\NP{k_r}}}{\pRe{\NP{k_r}}} 
\frac{\sqrt{E}}{k_r\NP{k_r^2-E}}}} \,\biggr ].
\end{equation}
In order to confirm our identification of the quantity in 
square brackets in \eqref{E2.5} with the 
density of states, we note as mentioned earlier, that 
(a) the energy derivative of the scattering phase shift, $d\delta_l/dE$, 
in the vicinity of a resonance, can be derived analytically by making use 
of the properties of the $S$-matrix and a theorem of Mittag-Leffler. 
For the case of an $s$-wave resonance, it is given by \cite{ourarxiv}, 
\begin{equation}\label{derivML}
\frac{d\delta_0(E)}{dE} \,= \, {\pIm{\frac{\sqrt{E}}{k_r\NP{k_r^2-E}}}} \, 
\end{equation}
and 
(b) the Beth and Uhlenbeck formula (\ref{densitystates}) allows us 
to relate the energy derivative of the phase shift with the density of states in an 
s-wave resonance as:
\begin{equation}\label{bethuhlenformula} 
\rho^{BU}_0(E) = \frac{1}{\pi} \, \frac{d\delta_0(E)}{dE} \, ,
\end{equation}
so that
\begin{equation}
\rho^{GC}_{iso}(E) = \rho^{BU}_0(E) + 
\, {\frac{1}{\pi} \pIm \,\biggl [ 
{ i\frac{\pIm{\NP{k_r}}}{\pRe{\NP{k_r}}} 
\frac{\sqrt{E}}{k_r\NP{k_r^2-E}}}} \,\biggr ].
\end{equation} 
The density of states as given by the Beth-Uhlenbeck formula is the 
same as the first term in \eqref{E2.7p}. 
The second term in \eqref{E2.7p} can be seen to be a small correction to 
the first term for narrow resonances. The reason for the 
correction term not appearing in the Beth-Uhlenbeck (BU) formula could be due to the 
approximations made in the derivation of the BU formula and remains to be investigated.
With the above confirmation, we 
conclude that the GC and DN formalisms (taken for the case of an isolated 
$s$-wave resonance),
and the Fock-Krylov method with the density given using the Beth-Uhlenbeck formula, 
are equivalent.

\section{Analytical expression for the survival amplitude}\label{S3}
Analytical expressions for the survival amplitude, $A(t)$, of a resonance given by a 
Breit-Wigner form for the energy density can be found in \cite{polaco,urbanowski2018}. 
In \cite{ourarxiv}, the analytical expressions for $A(t)$ were derived using generalized 
expressions for the energy density (derived using the analytical properties of the 
$S$-matrix and the Mittag-Leffler theorem) within the Fock-Krylov method.
The expressions were shown to reduce to those arising from the Breit-Wigner form alone 
plus corrections. 
Here, we shall find an analytical expression for the survival amplitude 
given in \eqref{E2.5} and (\ref{E26}), study their asymptotic behaviour and  
analyse the transition region between the exponential and non-exponential decay law.
Apart from obtaining analytical expressions for the transition time, we shall examine 
some nuclear and particle decays and the relevance of the results for an 
experimental observation of the non-exponential decay law.

\subsection{Survival Amplitude in terms of the incomplete gamma function}
In the present section, we shall provide analytical 
expressions for the survival amplitudes in a combined form which is valid for 
both methods. 
Analytical formulae for the survival amplitudes 
evaluated in \cite{Garcia1} within the Green's function 
method, were presented in terms of the error functions. 
Here we present the expressions using incomplete gamma functions. 

Substituting \eqref{density} in \eqref{E1.8}, the survival amplitude is given as 
\begin{eqnarray}\label{E3.1}
A(t)&=&\intseminf{\rho(E)e^{-iEt}}{E} \nonumber \\
&=&\frac{1}{2\pi i}\sum_p{\frac{\ga_p(k_p)}{k_p}\intseminf{\frac{\sqrt{E}}{k_p^2-E}\,e^{-iEt}}{E}}-
\frac{1}{2\pi i}\sum_p{\frac{\ga_p^*(k_p)}{k_p^*}\intseminf{\frac{\sqrt{E}}{{k_p^*}^2-E}\,e^{-iEt}}{E}} \, .\nonumber \\
\end{eqnarray}
Using \eqref{EB4} and \eqref{EB6} (see appendix \ref{B}), we get 
\begin{align}
A(t)
=&\frac{1}{2\pi i}\sum_p{\frac{\ga_p(k_p)}{k_p}\GC{2\pi ik_pe^{-ik_p^2t}+i\frac{\sqrt{\pi}}{2}k_pe^{-ik_p^2t}\Ga\NP{-\tfrac{1}{2},-ik_p^2t}}}\notag\\
&-\frac{1}{2\pi i}\sum_p{\frac{\ga_p^*(k_p)}{k_p^*}\GC{i\frac{\sqrt{\pi}}{2}k_p^*e^{-i{k_p^*}^2t}\Ga\NP{-\tfrac{1}{2},-i{k_p^*}^2t}}}\notag\\
=&\sum_p{\ga_p(k_p)e^{-ik_p^2t}}
+\frac{1}{4\sqrt{\pi}}\sum_p{\MC{\ga_p(k_p)e^{-ik_p^2t}\Ga\NP{-\tfrac{1}{2},-ik_p^2t}-\ga_p^*(k_p)e^{-i{k_p^*}^2t}\Ga\NP{-\tfrac{1}{2},-i{k_p^*}^2t}}}.\label{E3.2}
\end{align}
In order to ensure that the survival amplitude at $t$ = 0 is unity, using $\Ga\NP{-\frac{1}{2},0}=-2\sqrt{\pi}$ together with the 
property \eqref{coef1} gives:
\begin{equation}\label{E3.3}
A(0)=\sum_p{\ga_p(k_p)}-\frac{1}{2}\sum_p{\NC{\ga_p(k_p)-\ga_p^*(k_p)}}=\pRe\sum_p{\ga_p(k_p)}=1.
\end{equation}
Using the properties of the incomplete gamma functions \cite{Lebedev}:
\begin{equation}\label{E3.4}
\Ga(\al+1,z)=\al\Ga(\al,z)+z^\al e^{-z}
\end{equation}
with $\al=1/2$ and $z=-ik_p^2t$, we can write \eqref{E3.2} as
\begin{align*}
A(t)
=&\sum_p{\ga_p(k_p)e^{-ik_p^2t}}
+\frac{1}{4\sqrt{\pi}}\sum_p{\ga_p(k_p)\GC{{2}\NP{-ik_p^2t}^{-1/2}-2e^{-ik_p^2t}\Ga\NP{-\tfrac{1}{2},-ik_p^2t}}}\\
&-\frac{1}{4\sqrt{\pi}}\sum_p{\ga_p^*(k_p)\GC{{2}\NP{-i{k_p^*}^2t}^{-1/2}-2e^{-ik_p^2t}\Ga\NP{-\tfrac{1}{2},-i{k_p^*}^2t}}}\\
=&\sum_p{\ga_p(k_p)e^{-ik_p^2t}}-\frac{1}{2\sqrt{\pi}}\sum_p{\MC{\ga_p(k_p)e^{-ik_p^2t}\Ga\NP{\tfrac{1}{2},-ik_p^2t}-\ga_p^*(k_p)e^{-i{k_p^*}^2t}\Ga\NP{\tfrac{1}{2},-i{k_p^*}^2t}}}\notag\\
&+\frac{1}{\sqrt{\pi t}}e^{3i\pi/4}\pIm{\sum_p{\frac{\ga_p(k_p)}{k_p}}}.
\end{align*}
The last term is zero due to the property \eqref{coef2}. Finally,
\begin{equation}\label{E3.5}
A(t)=\sum_p{\ga_p(k_p)e^{-ik_p^2t}}
-\frac{1}{2\sqrt{\pi}}\sum_p{\MC{\ga_p(k_p)e^{-ik_p^2t}\Ga\NP{\tfrac{1}{2},-ik_p^2t}-\ga_p^*(k_p)e^{-i{k_p^*}^2t}\Ga\NP{\tfrac{1}{2},-i{k_p^*}^2t}}}.
\end{equation}
The above expression is equivalent to Eq. (4.21) of Ref. \cite{Sym1} given in terms 
of the M functions.
For a given $p$, each term depends on the pole associated with the index $p$ and 
it is possible to define partial survival amplitudes for the pole $k_p$ as:
\begin{equation}\label{E3.6}
A_p(t)=\ga_p(k_p)e^{-ik_p^2t}-\frac{1}{2\sqrt{\pi}}\MC{\ga_p(k_p)e^{-ik_p^2t}\Ga\NP{\tfrac{1}{2},-ik_p^2t}-\ga_p^*(k_p)e^{-i{k_p^*}^2t}\Ga\NP{\tfrac{1}{2},-i{k_p^*}^2t}},
\end{equation}
such that the survival amplitude takes the simple form:
\begin{equation}\label{E3.7}
A(t)=\sum_p{A_p(t)}.
\end{equation}
\subsection{Behaviour at large times} 
Using the asymptotic expansion of the incomplete gamma function \cite{Copson2}:
\begin{equation}\label{E3.8}
e^z\Ga\NP{\al,z}\sim z^{\al-1}+(\al-1)z^{\al-2}+\cdots
\end{equation}
and ignoring exponential terms, we have, 
\begin{eqnarray}
A(t)&\sim&-\frac{1}{2\sqrt{\pi}}\sum_p{\ga_p(k_p)\GC{\frac{1}{k_p}(-it)^{-1/2}-
\frac{1}{2k_p^3}(-it)^{-3/2}}}\nonumber \\
&+&\frac{1}{2\sqrt{\pi}}\sum_p{\ga_p^*(k_p)\GC{\frac{1}{k_p^*}(-it)^{-1/2}-\frac{1}{2{k_p^*}^3}(-it)^{-3/2}}}\nonumber \\
&=&-\frac{1}{2\sqrt{\pi}}(-it)^{-1/2}\cdot2i\pIm{\sum_p{\frac{\ga_p(k_p)}{k_p}}}
+\frac{1}{4\sqrt{\pi}}(-it)^{-3/2}\cdot2i\pIm{\sum_p{\frac{\ga_p(k_p)}{k_p^3}}}.
\end{eqnarray}
The first term is zero due to the properties of the coefficients, $\ga_p(k_p)$, and
\begin{equation}\label{E3.9}
A(t)\sim-\frac{e^{\,i\pi/4}}{\sqrt{4\pi}}\pIm{\EP{\sum_p{\frac{\ga_p(k_p)}{k_p^3}}}}t^{-3/2}.
\end{equation}
The above result has also been obtained in appendix \ref{A} by computing $A(t)$ with the 
steepest descent method. In both cases, the results are consistent and the 
survival probability is proportional to $t^{-3}$ for large $t$.
We remind the reader that the above analysis has been performed for $S$-waves.  
\subsection{Survival amplitude for an isolated resonance}
If the system under analysis has only one resonant pole $k_r$, 
the expression deduced for the survival amplitude at any time $t$ as well as that 
for large times can be written in a simple form:
\begin{align}
A(t)&={\ga_r(k_r)e^{-ik_r^2t}}
-\frac{1}{2\sqrt{\pi}}{\MC{\ga_r(k_r)e^{-ik_r^2t}\Ga\NP{\tfrac{1}{2},-ik_r^2t}-\ga_r^*(k_r)e^{-i{k_r^*}^2t}\Ga\NP{\tfrac{1}{2},-i{k_r^*}^2t}}},\label{E3.21}\\
A(t)&\sim-\frac{e^{\,i\pi/4}}{\sqrt{4\pi}}\pIm{\EC{{\frac{\ga_r(k_r)}{k_r^3}}}}t^{-3/2}.\label{E3.22}
\end{align}
Substituting the expression \eqref{E3.26}, Eq. \eqref{E3.21} and \eqref{E3.22} can be alternatively written as
\begin{align}
A(t)&={\frac{k_r}{\pRe{\NP{k_r}}}\,e^{-ik_r^2t}}
-\frac{1}{2\sqrt{\pi}\pRe{\NP{k_r}}}{\MC{k_re^{-ik_r^2t}\Ga\NP{\tfrac{1}{2},-ik_r^2t}-k_r^*e^{-i{k_r^*}^2t}\Ga\NP{\tfrac{1}{2},-i{k_r^*}^2t}}},\label{E3.27}\\
A(t)&\sim-\frac{e^{\,i\pi/4}}{\sqrt{4\pi}\pRe{\NP{k_r}}}\pIm{\EP{{\frac{1}{k_r^2}}}}t^{-3/2}.\label{E3.28}
\end{align}
An inspection of \eqref{E3.27} reveals that, for intermediate times, 
the survival amplitude can be described by an exponential function, i.e., 
\begin{equation}\label{E3.11}
A_r(t)\approx \frac{k_r}{\pRe{\NP{k_r}}}\,e^{-ik_r^2t}.
\end{equation}

The quantum mechanical description of the decay law leads to a non-exponential behaviour 
at very short and very large times with the intermediate region being dominated by the 
exponential decay law. In what follows, we shall concentrate on the transition region 
from the exponential to the power law at large times. 
Considering the case of an isolated resonance, 
the critical time for the transition to the power law is investigated and its 
relevance for an experimental observation of the power law is discussed.

\section{Critical time}\label{S4.3}
It would be useful if we could find the parameters on 
which the critical time for the survival amplitude to go 
from an exponential to a power law behaviour depends. 
With this objective, we shall study the intersection of the 
intermediate and large time survival probabilities. 
We define the {\it critical time} between these behaviours as $t_c$, such that 
\begin{equation}\label{E3.12}
\GB{\frac{k_r}{\pRe{\NP{k_r}}}\,e^{-ik_r^2t_c}}^2=\GB{-\frac{e^{\,i\pi/4}}{\sqrt{4\pi}\pRe{\NP{k_r}}}\pIm{\GP{{\frac{1}{k_r^2}}}}t_c^{-3/2}}^2.
\end{equation}
Since $k_r^2=\ep_r-i\Ga_r/2$ and defining $\tau_c=\Ga_rt_c$, 
it is convenient to write \eqref{E3.12} as
\begin{equation}\label{E3.13}
\NB{k_r}^2e^{-\tau_c}=\frac{\Ga_r^3}{{4\pi}}\GB{\pIm{\GP{{\frac{1}{k_r^2}}}}}^2\tau_c^{-3}.
\end{equation}
Let $C$ be the constant defined by
\begin{equation}\label{E3.14}
C=\frac{\Ga_r^3}{{4\pi}}\GB{\frac{1}{k_r}\pIm{\GP{{\frac{1}{k_r^2}}}}}^2,
\end{equation}
which is always positive. The transition time, $\tau_c = \Gamma_r t_c$, 
shall be the zero of the function
\begin{equation}\label{E3.15}
f(\tau_c)=e^{-\tau_c}-C\tau_c^{-3}=\frac{\tau_c^3e^{-\tau_c}-C}{\tau_c^3}.
\end{equation}
Let the auxiliary function $g(\tau_c)$ together with its derivative be given by 
\begin{align}
g(\tau_c)&=\tau_c^3e^{-\tau_c}-C,\label{E3.16}\\
g'(\tau_c)&=\tau_c^2(3-\tau_c)e^{-\tau_c}.\label{E3.17}
\end{align}
$g(\tau_c)$ has two critical points: $\tau_c=0$ and $\tau_c=3$. 
In the interval $0<\tau_c<3$, $g'(\tau_c)>0$; 
and in the interval $\tau_c>3$, $g'(\tau_c)<0$. This means that $\tau_c=3$ is a 
maximum and $\tau_c=0$ is a minimum. The values of $g$ at those points 
are $g(0)=-C$ and $g(3)=27e^{-3}-C$. When $\tau_c\to\infty$, $g\to-C$.

Since $C>0$, $g$ is positive on some interval or negative for all $\tau_c>0$, 
and this depends on the sign of its maximum. If $27e^{-3}-C<0$, 
the maximum is negative and $g<0$ in $\tau_c>0$. If $C=27e^{-3}$, 
the maximum is zero and $g\leq0$ in $\tau_c>0$; but, if $27e^{-3}-C>0$, 
$g$ will have two zeros and will be positive in the interval formed by those zeros.

Now, it is easy to find the zeros of $f$ and this depends on the 
values of $C$. We have three cases:
\begin{enumerate}[i)]
\item First case: If $C>27e^{-3}$, $f(\tau_c)$ has no zeros and is negative 
for $\tau_c>0$, this means that $e^{-\tau_c}<C\tau_c^{-3}$: 
there is no critical time and thus, the power law behaviour always dominates. 
\item Second case: If $C=27e^{-3}$, $f(\tau_c)$ has one zero and is negative or 
null in $\tau_c>0$, this implies $e^{-\tau_c}\leq C\tau_c^{-3}$: 
there is only one critical point and the power law behaviour dominates again. 
\item Third case: If $C<27e^{-3}$, $f(\tau_c)$ has two zeros 
$\tau_{c1}$ and $\tau_{c2}$ such that $\tau_{c1}<\tau_{c2}$. 
$f(\tau_c)>0$ for $\tau_{c1}<\tau_c<\tau_{c2}$ (and $e^{-\tau_c}>C\tau_c^{-3}$ here). 
$f(\tau_c)<0$ for values of $\tau_c$ out of this interval and  
$e^{-\tau_c}<C\tau_c^{-3}$: the exponential behaviour is more dominant than the power 
law behaviour in $\tau_{c1}<\tau<\tau_{c2}$, but, for $\tau>\tau_{c2}$, it is the 
power law that dominates. We shall see that $\tau_{c2}$ can be identified
as the critical time for the transition from the exponential to the power law. 
\end{enumerate} 

Let us now see if it is possible to write the parameter $C$ in terms of 
$x_r=\dfrac{\Ga_r}{2\ep_r}$. 
For a given resonance pole, $E_r - i\Gamma_r/2$ in the complex energy plane, 
$\epsilon_r$ is defined as $E_r - E_{th}$ where $E_{th}$ for example 
is the sum of the masses of the decay products of an unstable particle with 
mass $E_r$. Since 
\[
\frac{1}{k_r^2}=\GB{\frac{1}{k_r^2}}\exp{\NC{-i\pArg{\NP{k_r^2}}}},
\]
we have 
\[
\pIm\GP{\frac{1}{k_r^2}}=-\GB{\frac{1}{k_r^2}}\sin{\NC{\pArg{\NP{k_r^2}}}},
\]
and $C$ is equal to
\[
C=\frac{2}{\pi}\GP{\frac{\Ga_r}{2}}^3\,\frac{1}{|k_r^2|^3}\sin^2{\NC{\pArg{\NP{k_r^2}}}}.
\]
But, $|k_r^2|=\ep_r\sqrt{1+x_r^2}$, and $\sin{\NC{\pArg{\NP{k_r^2}}}}=-x_r/\sqrt{1+x_r}$. Thus,
\begin{equation}\label{E3.18}
C=\frac{2}{\pi}\,\frac{x_r^5}{\NP{1+x_r^2}^{5/2}}=\frac{2}{\pi}\GP{\frac{x_r}{\sqrt{1+x_r^2}}}^5.
\end{equation}
Thus $C$ can be written as a function of $x_r$.  
An upper bound of $C$ can be obtained if we see that the term 
$x_r\NP{1+x_r^2}^{-1/2}$ is always less than one for any value of $x_r$. Thus,
\begin{equation}\label{E3.19}
C<\frac{2}{\pi}.
\end{equation}
This bound is less than $27e^{-3}=1.34\dotsc$ and the third case applies always.  

Coming back to the definition of $\tau_{c}$ through the zeros of the 
function $f(\tau_c)$ in Eq. \eqref{E3.15}, we can write it as
\[
-\frac{\tau}{3}e^{-\tau/3}=-\frac{\sqrt[3]{C}}{3}.
\]
With a change of variables, $z=-\tau/3$, we can write
\begin{equation}\label{tausolution}
ze^z=-\frac{\sqrt[3]{C}}{3}. 
\end{equation}
Here we note that the Lambert function $W(x)$ is the inverse function of the 
function $x=W(x)e^{W(x)}$. Although this function has infinite branches, 
we would be interested in the real branches: the principal one, denoted by $W_0(x)$,  
which takes the values $W_0(x)\geq-1$ for $x\geq-1$; and the second one $W_{-1}(x)$, 
which takes the values $W_{-1}(x)\leq-1$ for $-1\leq x\leq0$. 
Thus the solution of Eq. \eqref{tausolution} would be, 
according the definition of the Lambert function,
\[
z=-\frac{\tau}{3}=W{\GP{-\frac{\sqrt[3]{C}}{3}}}.
\]
In order to decide which branch, we recall that $0<C<\dfrac{2}{\pi}$. Thus, 
the argument of the Lambert function satisfies 
\[
-\dfrac{1}{3}\sqrt[3]{\dfrac{2}{\pi}}<0,
\]
or $-0.2867513379<C<0$. If we use the branch $W_0(x)$, we will obtain a critical 
time that satisfies $0<\tau<1.3484499515$: this interval corresponds to the solutions 
$\tau_{c1}$, the smaller times, which we  discard. However, 
if we use the other branch, the critical time satisfies $\tau>5.6426374987$ and 
we associate it with the transition time $\tau_c$. 
The appropriate solution is given by, 
\begin{equation}\label{taulambert} 
\tau_c=-3W_{-1}{\GP{-\frac{\sqrt[3]{C}}{3}}}.
\end{equation}

The above formula for $\tau_c$ is model independent if the energy density $\rho(E)$ 
entering the calculation of the survival amplitude is model 
independent. This density as mentioned above, is up to a factor, the same as the one 
obtained in \cite{ourarxiv} solely using the properties of the $S$ matrix and a 
theorem of Mittag-Leffler. In \cite{bogda}, the authors had also obtained an equation 
similar to Eq. \eqref{tausolution} but for a Breit-Wigner form of $\rho(E)$
without a threshold factor. 
The solution of the equation in \cite{bogda} can be written as
\begin{equation}\label{taulambertbw} 
\tau_c^{BW}=-W_{-1}{\GP{-\frac{4}{\pi} x_r^2}}.
\end{equation}
In Table I, values of $\tau_{c}$ (this work) and $\tau_{c}^{BW}$ (as in 
\cite{bogda}) for some values of $x_r$ are compared. The absence of the 
threshold factor (apart from the use of a Breit-Wigner form) 
gives rise to smaller critical times $\tau_c^{BW}$ as compared to 
$\tau_c$ evaluated from the model independent form involving the 
correct threshold for $\rho(E)$ as in \eqref{derivML}. The second 
column displays some fitted values, $\tau_{c}^{fit}$, to be discussed below.

\begin{table}\label{tab1}
\caption{Critical values $\tau_c = \Gamma_r t_c$ of the transition to the 
non-exponential power law behaviour of the survival probability as a 
function of the parameter $x_r = \Gamma_r/ 2 \epsilon_r$ in the model 
independent (\ref{taulambert}), Breit-Wigner parametrization (\ref{taulambertbw}) 
and fitted parametrization (\ref{tauGC}) cases.}
\begin{tabular}{cccccc}
\begin{tabular}{|c|c|c|c|}\hline
$x_r$  & $\tau_{c} \eqref{taulambert}$ & $\tau_{c}^{fit}$ \eqref{tauGC}&
$\tau_{c}^{BW}$ \eqref{taulambertbw} 
\\\hline 
0.1    & 21.1& 21 & 15.6 \\\hline
0.2    & 17.1& 17.2 & 12.4 \\\hline
0.3    & 14.8& 15 & 10.4 \\\hline
0.4    & 13 & 13.5  & 9 \\\hline
0.5    & 11.9& 12.3 & 7.8 \\\hline
\end{tabular} &$\quad$&
\begin{tabular}{|c|c|c|c|}\hline
$x_r$  & $\tau_{c}$ \eqref{taulambert} & $\tau_{c}^{fit}$ \eqref{tauGC} 
&$\tau_{c}^{BW}$ \eqref{taulambertbw}
\\\hline
0.6 & 11&11.3  & 6.8 \\\hline
0.7 & 10.2 &10.4 & 5.9 \\\hline
0.8 & 9.6&9.7 & 5.0 \\\hline
0.9 & 9.1&9.1 & 4.2 \\\hline
1   & 8.7&8.5  & 3.3 \\\hline
\end{tabular} &$\quad$&
\end{tabular}
\end{table}
Though, in principle, 
most real life resonances would correspond to $x_r < 1$, it is interesting to 
note that for $C=2/\pi$ (or $x_r\to\infty$), 
we have the lower bound of the critical time: $\tau_{c}=5.6426375$. 

\begin{figure}[ht]
\centering
\includegraphics[scale=0.7]{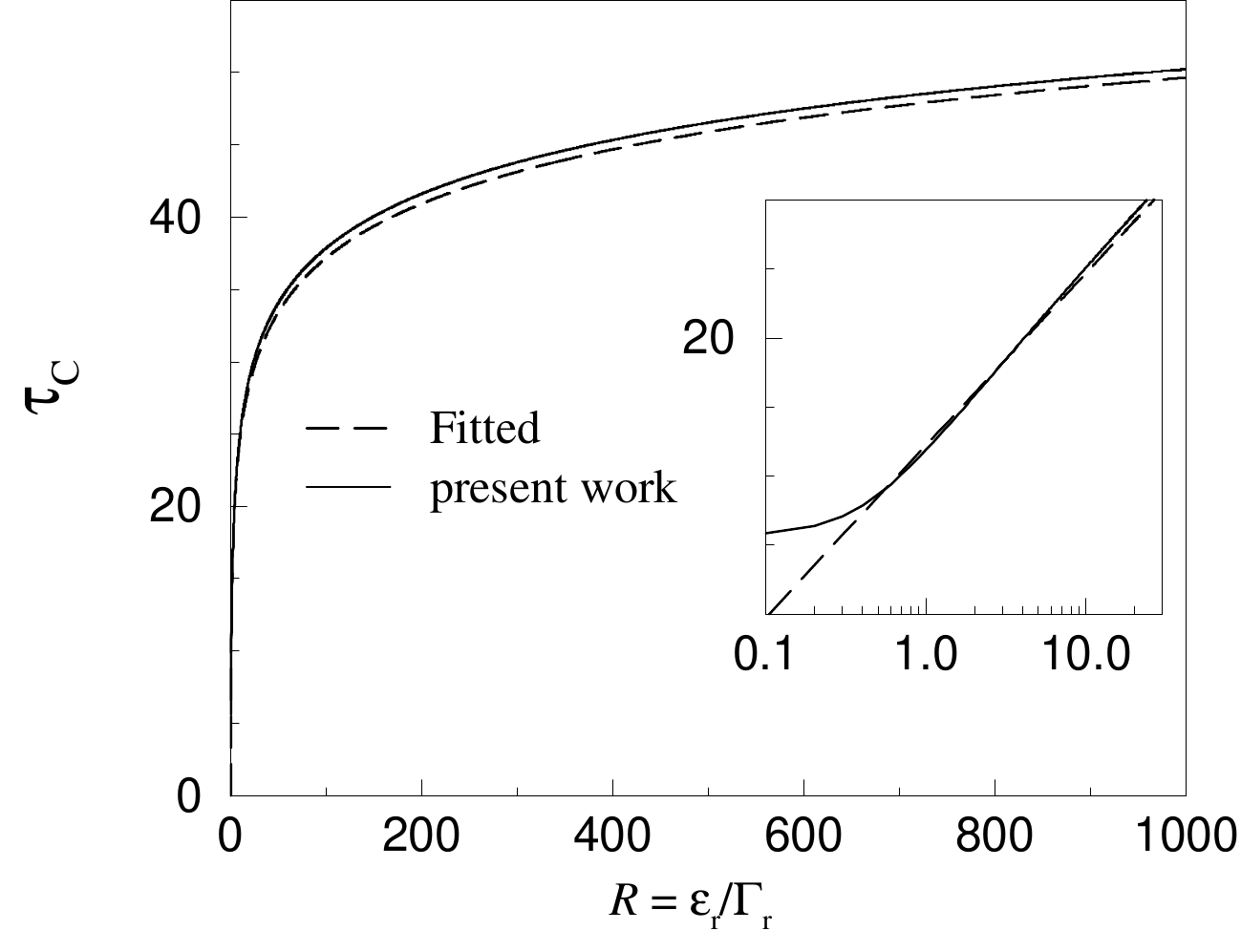}
\caption{Comparison of critical times as a function of the variable 
$R = \epsilon_r/\Gamma_r$ using Eqs (\ref{taulambert}) and 
(\ref{tauGC}). The inset shows the same figure for a smaller range of $R$.}
\label{criticalts}
\end{figure}

The critical time for the transition from the exponential to the power law at large 
times was also studied in \cite{garciaisolated} in the context of a single isolated 
resonance. Determining the transition time from a numerical calculation of the 
survival probabilities for several values of the variable 
$R = \epsilon_r/\Gamma_r$ 
and observing its behaviour as a function of this variable, the authors 
assumed a logarithmic form for the transition time as follows:
\begin{equation}\label{tauGC}
\tau_{c}^{fit} = A \ln{(R)} + B\, 
\end{equation}
and obtained the values $A$= 5.41 and $B$ = 12.25 from a fitting procedure. 
This formula is a refinement of the estimate, $\tau_L = K \ln{(R)}$ of
Winter \cite{winter1962}. 
In 
Fig. \ref{criticalts}, we show a comparison of $\tau_{c}^{fit}$ and $\tau_{c}$ evaluated 
from the analytical expression (\ref{taulambert}) as a function of the variable $R$. 
It must be noted that (i) even though $\tau_{c}$ and $\tau_{c}^{fit}$ are very similar 
for most values of $R$, for small $R$ which corresponds to the case of broad resonances 
(such as the sigma meson for example \cite{pdg}) they can be quite different and (ii) 
whereas $\tau_{c}$ has a finite lower limit of about 5.64 mentioned above, 
$\tau_{c}^{fit}$ can even take negative values for very small $R$.
The inset in 
Fig. \ref{criticalts} shows that below $R$ = 0.3, the 
analytical expression (\ref{taulambert}) and the fitted one, 
Eq. (\ref{tauGC}) start differing. 
The region of $R < 1$ has indeed been found to be important in literature 
in connection with the decay of artificial quantum structures \cite{Garcia2006}. In \cite{Garcia2006}, the authors found that the decay law could have a 
non-exponential form at all times, for the range, 0 $<$ $R \leq$ 0.3. 
The decay of a single ultracold atom was also shown to be non-exponential 
in \cite{Garcia2016} below $R$ = 0.3. The findings of \cite{Garcia2006} and 
\cite{Garcia2016} essentially imply that there is no transition time. 

\section{Non-exponential decay of particles and nuclei}\label{s6}
We shall now apply the results obtained in this work to study some unstable states 
which have been investigated experimentally. In Table II, we list the critical times 
for the beginning of the non-exponential (power) law for the particles and nuclei 
which have been studied in literature \cite{norman,rutherford,buttwilson,nikolaev}.
The transition time as calculated in the present work appears 
many half-lives later than the number of observed half-lives. 
From the values given in the table,  
it is evident that (a) it was necessary to wait much longer to observe the power law 
(b) but waiting so long would also destroy most of the sample with nothing left for 
measurement. 
One could then think of observing the broader resonances such as the sigma meson 
with a width of a few hundred MeV leading to a very small $\tau_c \sim 9 $, 
however, 
such a width corresponds to a lifetime of about 10$^{-23}$ s, making the observation 
once again not possible.  
\begin{table}\label{tab2}
\caption{Critical values $\tau_c = \Gamma_r t_c$ of the transition to the 
non-exponential behaviour for experimentally measured particles and nuclei.}
\begin{tabular}{|c|c|c|c|c|}\hline
    &Lifetime&$x_r = \Gamma_r/2\epsilon_r$  &$\tau_{c} \eqref{taulambert}$ &Number of half-lives \\
    &  & &  & measured\\\hline
$^{56}$Mn(3$^+$)$\to \,^{56}$Fe(2$^+$) + e$^-$ + $\bar{\nu}_e$  \cite{norman}   
&2.5789 h & 1.2 $\times$ 10$^{-26}$  &316  & 45  \\\hline
$^{222}$Rn $\to\, ^{218}$Po + $\alpha$ \cite{rutherford,buttwilson} &3.8235 d &1.2 
$\times$ 10$^{-28}$  &339  & 27, 40  \\\hline
 $K^+ \to \mu^+\, \nu_{\mu}$ \cite{nikolaev}  &12.443 ns & 4.5 $\times$ 10$^{-17}$  &204  & 7.3 \\\hline
 $K^+ \to \pi^+ \, \pi^0$ \cite{nikolaev}  &12.265 ns & 8.4 $\times$ 10$^{-17}$ 
&201  & 4 \\\hline
\end{tabular}
\end{table}
\section{Interference region}\label{S4.4}
In one of the early works on the time evolution of unstable states, 
the oscillatory character of the survival probability at short and large 
times was demonstrated by Winter in a barrier penetration problem \cite{winter}.
For narrow resonances, i.e., for 
$x_r = \Gamma_r/2 \epsilon_r \ll 1$, following the 
non-exponential behaviour at very short times, the survival probability displays  
a prominent exponential decay law followed by a strong oscillatory transition 
region (several half-lives $\tau_C = \Gamma_r t_c$ later) which is then 
followed by the power law at large times. 
The origin of this particular oscillation lies in 
the interference of the exponential and power law behaviours.
In this section, we 
shall investigate the large time transition region and obtain an 
analytical expression to describe it.  
\subsection{Origin of the oscillatory term}\label{S4.4.1} 
\begin{figure}[!htb]
\centering
\includegraphics[scale=1.1]{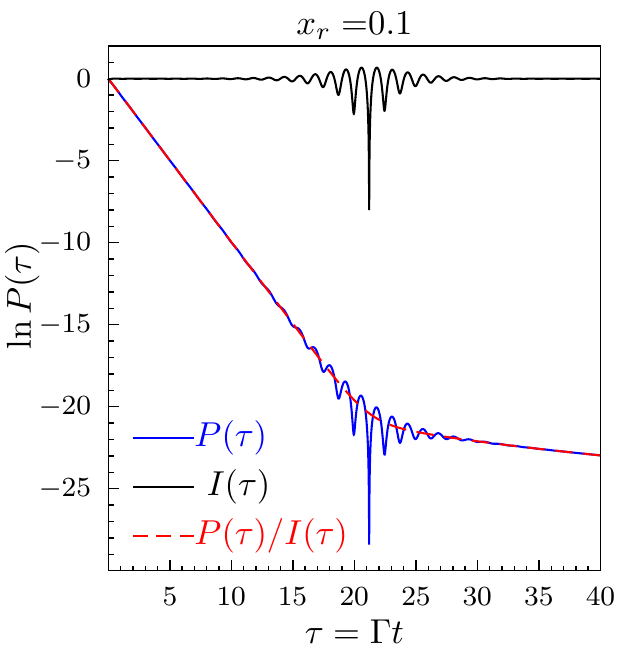}
\caption{Survival probability (on a logarithmic scale). The blue curve 
is the full survival probability and the red line displays the behaviour 
without the modulating function (shown in black at the top). The ratio
$x_r = \Gamma_r /2 \epsilon_r$ here is chosen to be 0.1.}\label{F4.4.1a}
\end{figure}

\begin{figure}[!htb]
\centering
\includegraphics[scale=1.3]{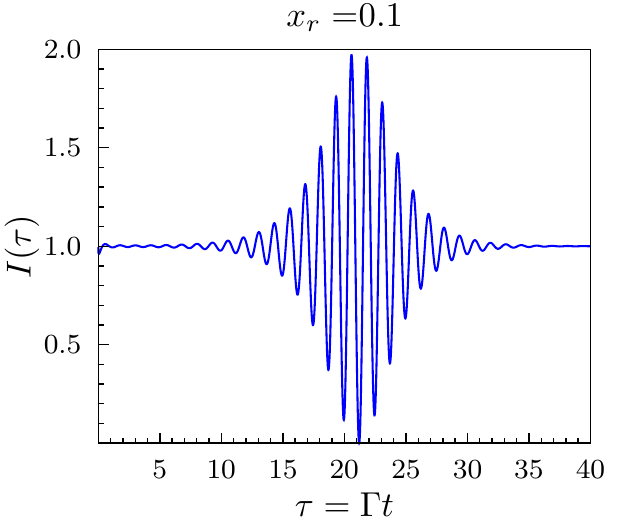}
\caption{Modulating function shown in Fig. \ref{F4.4.1a}, now on a linear scale.}
\label{F4.4.1b}
\end{figure}

As we have already seen, the survival amplitude can be expressed as a sum of two parts: 
one describing an exponential decay, $A_e$ and another term $A_p$ with a power law 
behaviour. Thus, the total amplitude $A(t)$ is given by, 
\begin{equation}\label{E4.4.1}
A(t)=A_e(t)+A_p(t),
\end{equation}
with 
\begin{align}
A_e(t)=&{\frac{k_r}{\pRe{\NP{k_r}}}\,e^{-ik_r^2t}},\label{E4.4.2}\\
A_p(t)=&-\frac{1}{2\sqrt{\pi}\pRe{\NP{k_r}}}{\MC{k_re^{-ik_r^2t}\Ga\NP{\tfrac{1}{2},-ik_r^2t}-k_r^*e^{-i{k_r^*}^2t}\Ga\NP{\tfrac{1}{2},-i{k_r^*}^2t}}}\notag\\
=&-\frac{e^{\,i\pi/4}}{\sqrt{4\pi}\pRe{\NP{k_r}}}\pIm{\EP{{\frac{1}{k_r^2}}}}t^{-3/2}+O(t^{-5/2}),\quad t\to\infty,\label{E4.4.3}
\end{align}
where $k_r^2=\ep_r-i\Ga_r/2$. The survival probability is 
\begin{align}
P(t)=|A(t)|^2&=|A_e(t)|^2+|A_p(t)|^2+2\pRe{\NC{A_e(t)A_p^*(t)}}\notag\\
             &=P_e(t)+P_p(t)+2\pRe{\NC{A_e(t)A_p^*(t)}},\label{E4.4.4}
\end{align}
where $P_e(t)=|A_e(t)|^2$ and $P_p(t)=|A_p(t)|^2$. 
Given the fact that the oscillatory behaviour becomes evident on a logarithmic scale, 
we rewrite the above equation as, 
\begin{equation}\label{E4.4.5}
P(t)=\MC{P_e(t)+P_p(t)}\EL{1+\frac{2\pRe{\NC{A_e(t)A_p^*(t)}}}{|A_e(t)|^2+|A_p(t)|^2}}=
I(t)\MC{P_e(t)+P_p(t)},
\end{equation}
where we have defined a modulating function, $I(t)$, such that 
\begin{equation}\label{E4.4.7}
I(t)=1+\frac{2\pRe{\NC{A_e(t)A_p^*(t)}}}{|A_e(t)|^2+|A_p(t)|^2}.
\end{equation}
Taking the logarithm on both sides of the above equation, we can write, 
\begin{equation}\label{E4.4.6}
\ln{P(t)}=\ln{\MP{P_e(t)+P_p(t)}}+\ln{I(t)}.
\end{equation}
Eq. \eqref{E4.4.6} hints that the modulating function $I(t)$ 
must give rise to the oscillations and this is indeed confirmed in Fig. \ref{F4.4.1a}.
The modulating function $I(t)$ is shown in Fig. \ref{F4.4.1b} on a linear scale. 

\subsection{Analytical expression for the modulating function}\label{S4.4.2}
\begin{figure}[!htb]
\centering
\includegraphics[scale=1.2]{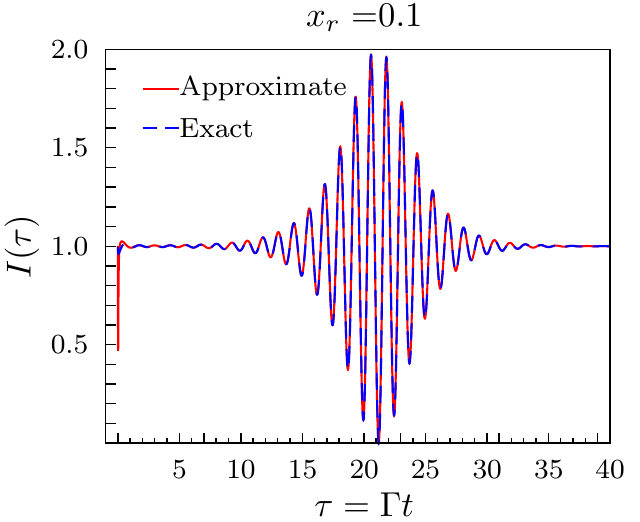}
\caption{Approximate and exact form of the modulating function.}\label{F4.4.3}
\end{figure}

If we naively replace Eqs 
\eqref{E4.4.2} and \eqref{E4.4.3} (first line) 
in Eq. \eqref{E4.4.7}, the modulating function 
can be written as, 
\begin{align}
I(t)&=1+\cfrac{2\pRe{\NC{A_e(t)A_p^*(t)}}}{|A_e(t)|^2+|A_p(t)|^2}\notag\\
&=1-\frac{1}{\sqrt{\pi}}\pRe{\frac{{\NC{\ga_r(k_r)e^{-ik_r^2t}}\NC{\ga_r(k_r)e^{-ik_r^2t}\Ga\NP{\tfrac{1}{2},-ik_r^2t}-\ga_r^*(k_r)e^{-i{k_r^*}^2t}\Ga\NP{\tfrac{1}{2},-i{k_r^*}^2t}}^*}}{\MB{\ga_r(k_r)e^{-ik_r^2t}}^2+\cfrac{1}{4{\pi}}\MB{{\ga_r(k_r)e^{-ik_r^2t}\Ga\NP{\tfrac{1}{2},-ik_r^2t}-\ga_r^*(k_r)e^{-i{k_r^*}^2t}\Ga\NP{\tfrac{1}{2},-i{k_r^*}^2t}}}^2}},\label{E4.4.8}
\end{align}
where $\ga_r(k_r)$ is given by eq. \eqref{E3.26}. Thus, the above equation as such would be quite difficult to analyze and hence we 
consider approximating $A_p(t)$ simply by the power law behaviour at large times.
Such an approximation is quite good for small values of $x_r$ where the critical time 
for the transition from the exponential to the power law behaviour 
(as seen in an earlier section) is quite large. 
Thus the expressions which will be derived below, will be valid only 
for resonances where $x_r\ll1$. 
With $\tau=\Ga_rt$ and $k_r^2=\ep_r-i\Ga_r/2$, we now write, 
\begin{equation}\label{E4.4.9}
-ik_r^2t=-\frac{1}{2}\tau-i\om_r\tau,
\end{equation}
where $\om_r$ is defined as 
\begin{equation}\label{E4.4.10}
\om_r=\frac{\ep_r}{\Ga_r}.
\end{equation}
The expressions \eqref{E4.4.2} and \eqref{E4.4.3} can now be written as, 
\begin{align}
A_e(t)&=\frac{k_r}{\pRe{\NP{k_r}}}\,e^{-ik_p^2t}=\frac{k_r}{\pRe{\NP{k_r}}}\,e^{-\tau/2}\,e^{-i\om_r\tau},\label{E4.4.11}\\
A_p(t)&=-\frac{e^{\,i\pi/4}}{\sqrt{4\pi}\pRe{\NP{k_r}}}\pIm{\GP{{\frac{1}{k_r^2}}}}t^{-3/2}=
-\frac{e^{\,i\pi/4}}{\pRe{\NP{k_r}}}\,\sqrt{\frac{\Ga_r^3}{4\pi}}\,\pIm{\GP{{\frac{1}{k_r^2}}}}\tau^{-3/2},\label{E4.4.12}
\end{align}
and the modulating function becomes 
\begin{align}
I(t)
&=1+\frac{2\pRe{\NC{A_e(t)A_p^*(t)}}}{|A_e(t)|^2+|A_p(t)|^2}\notag\\
&=1-\EC{2\NB{k_r}\sqrt{\frac{\Ga_r^3}{4\pi}}\pIm{\GP{{\frac{1}{k_r^2}}}}}\,\cfrac{e^{-\tau/2}\tau^{-3/2}\pRe{\NP{e^{-i\om_r\tau}\,e^{-i\pi/4}\,e^{\,i\pArg{k_r}}}}}{\NB{k_r}^2\,e^{-\tau}+\cfrac{\Ga_r^3}{4\pi}\,\GB{\pIm{\GP{{\cfrac{1}{k_r^2}}}}}^2\tau^{-3}}.\label{E4.4.13}
\end{align}
Introducing the constant $C$ given by Eq. \eqref{E3.14}, we get, 
\begin{equation}\label{E4.4.14}
I(\tau)=1+D\EP{\frac{e^{-\tau/2}\tau^{-3/2}}{e^{-\tau}+C\tau^{-3}}}\,\cos{\NP{\om_r\tau+\pi/4-\pArg{k_r}}},
\end{equation}
where $D$ is given by, 
\begin{equation}\label{E4.4.15}
D=-\frac{2}{\NB{k_r}}\sqrt{\frac{\Ga_r^3}{4\pi}}\pIm{\GP{{\frac{1}{k_r^2}}}}.
\end{equation}
{The modulating function so derived allows us to infer that:}
\begin{enumerate}[i)]
\item $I(\tau)$ oscillates about $I=1$ with a frequency $\om_r$.
\item The function is modulated with an amplitude 
\begin{equation}\label{E4.4.16}
m(\tau)=\frac{e^{-\tau/2}\tau^{-3/2}}{e^{-\tau}+C\tau^{-3}}.
\end{equation}
which is expected to be maximum at the critical time.
\item Apart from the above, the function $I(\tau)$ 
is expected to present problems for small values of $\tau$ (since 
we approximated $A_p(t)$ by its behaviour at large times).
\item Since $\om_r={1}/{2x_r}$, $I(\tau)$ is expected to oscillate a lot if 
$x_r$ is small. This will not be the case for $x_r$ close to or bigger than unity 
(see for example the case of the broad $\sigma$ meson where one observes no 
oscillation at all \cite{Kelkar1}).
\end{enumerate}
In Fig. \ref{F4.4.3}, we compare the modulating function calculated using  
Eq. \eqref{E4.4.8} (with  the complete analytical expressions for $A_p(t)$) 
and that using the approximation of the large time behaviour mentioned above. 
As expected, the function presents problems at small times but the approximation 
of using the large time behaviour instead of the exact expression is quite good.
Before resolving the problem at small times, let us first study the function $m(\tau)$.
\subsection{Analysis of \texorpdfstring{$m(\tau)$}{}}
\begin{figure}[!htb]
\centering
\includegraphics[scale=1.2]{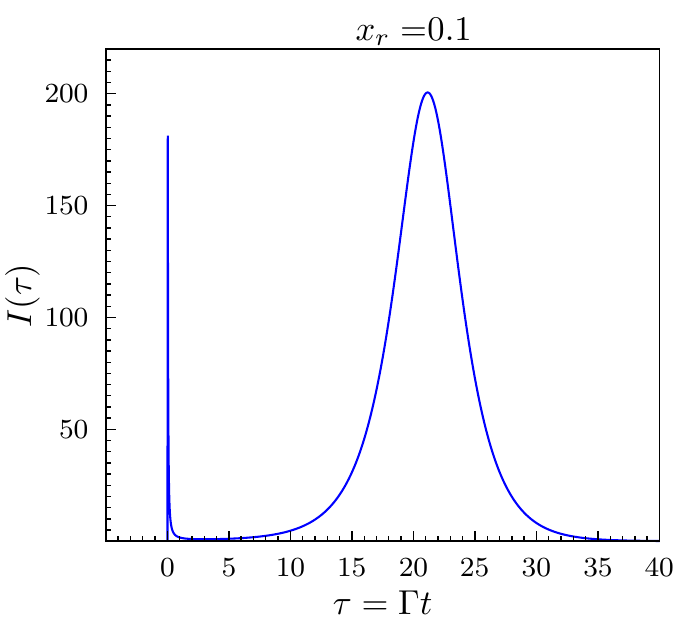}
\caption{$m(\tau)$ given by Eq. \eqref{E4.4.16}}\label{F4.4.4}
\end{figure}

If we write the function as follows: 
\[
m(\tau)=\frac{1}{e^{-\tau/2}\tau^{3/2}+Ce^{\,x/2}\tau^{-3/2}},
\]
then its derivative is given as 
\[
m'(\tau)=\frac{1}{2}\,\frac{^{1/2}e^{\,\tau/2}(\tau-3)(\tau^3-Ce^{\,\tau})}{(e^{-\tau/2}\tau^{3/2}+Ce^{\,x/2}\tau^{-3/2})^2}.
\]
The critical times in this function are 
$\tau=0,\tau_{c1},3,\tau_{c2}$ (in increasing order), 
where the second and fourth ones are solutions of 
$\tau^3-Ce^{\,\tau}=0$ and as analyzed in section \ref{S4.3} has two real solutions. 
It is easy to see that $\tau=0$ and $\tau=3$ are minima, while 
$\tau=\tau_{c1},\tau_{c2}$ are maxima.  
In Fig. \ref{F4.4.4} we show $m(\tau)$ for $x_r=0.1$. 
Here, $\tau_{c1}=0.0184942$ and $\tau_{c2}=21.143362$ 
with the latter corresponding to the critical time for the transition from the 
exponential to the power law behaviour (see Table I). 
The most relevant observation here is that 
$m(\tau)$ does display a maximum at the critical time as expected. 
However, in order to have an $m(\tau)$ that describes the modulating function 
correctly, we must get rid of the maximum close to $\tau=0$. 
One way of doing this could be by constructing a function of $\tau-\tau_c$ such that 
at $\tau=\tau_c$, it is given by $m(\tau_c)$, which is $\dfrac{1}{2\sqrt{C}}$.

\begin{figure}[!htb]
\centering
\includegraphics[scale=1.2]{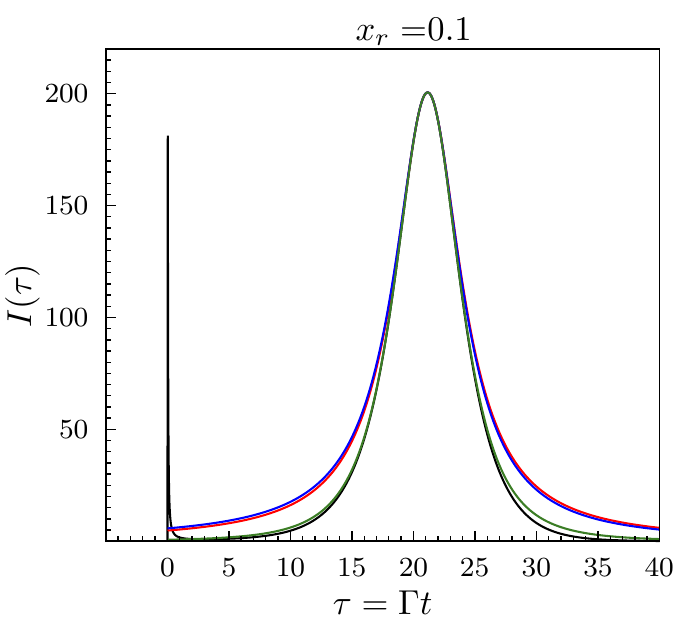}
\caption{$m(\tau)$ with the corresponding correction terms given in 
\eqref{E4.4.17}: blue line (second order), violet (third order) 
and green line (fourth order).}\label{F4.4.5}
\end{figure}

\begin{figure}[!htb]
\centering
\includegraphics[scale=1.2]{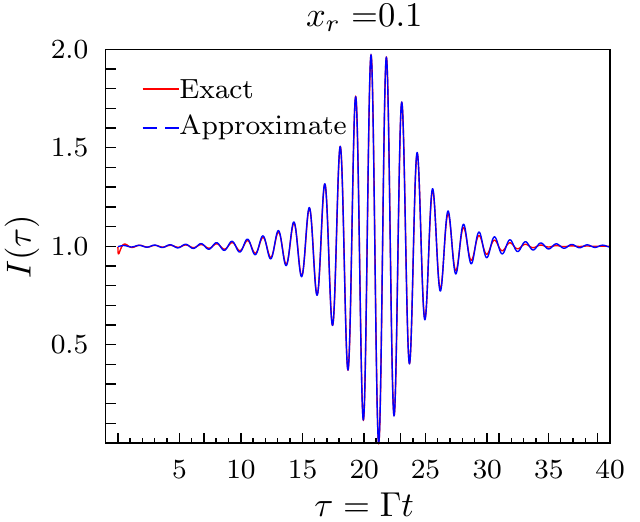}
\caption{Exact modulating function (red line) and the approximate one 
(blue line).}\label{F4.4.6}
\end{figure}

\begin{figure}[!htb]
\centering
\includegraphics[scale=1.2]{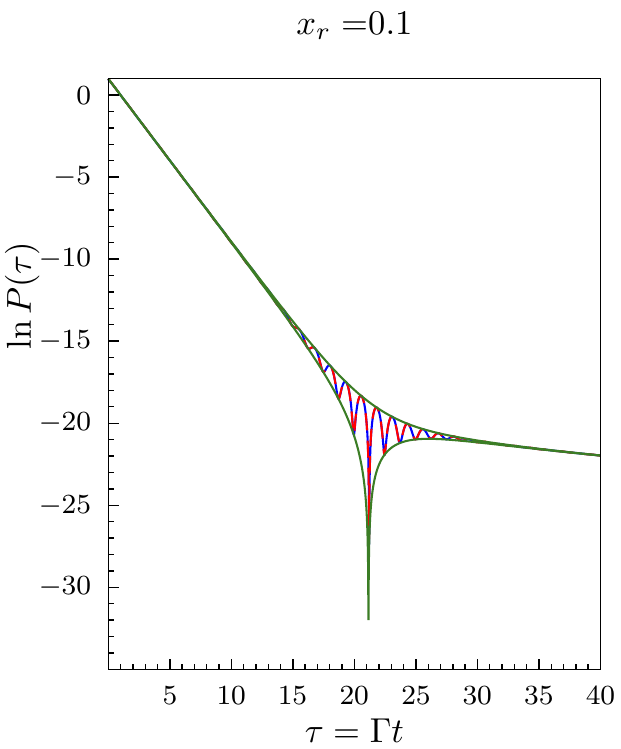}
\caption{Survival probability calculated exactly (blue line) and 
calculated using the approximate modulating function (red line). 
Green lines show the curves evaluated without the oscillatory part in 
the modulating function (see text).}\label{F4.4.7}
\end{figure}

The best way to do this is by expanding 
$\dfrac{1}{m(\tau)}$ in a series of $\tau-\tau_c$, so that, 
\begin{align}
\frac{1}{\sqrt{C}\,m(\tau)}
&=\sum_{n=0}^{\infty}{\frac{\NP{\tau-\tau_c}^n}{2^nn!}
\sum_{s=0}^{n}{\binom{n}{k}\EC{\frac{\Ga\NP{-\tfrac{1}{2}}}{\Ga\NP{-\tfrac{1}{2}-k}}+(-1)^{n+k}\frac{\Ga\NP{\tfrac{5}{2}}}{\Ga\NP{\tfrac{5}{2}-k}}}\GP{\frac{2}{\tau_c}}^k}}\notag\\
&=2+\frac{1}{4}\GP{1-\frac{3}{\tau_c}}^2\NP{\tau-\tau_c}^2
+\frac{1}{48}\GP{\frac{36}{\tau_c^2}-\frac{108}{\tau_c^3}}\NP{\tau-\tau_c}^3\notag\\
&\ \ \ \ \ \ \ \ \ \ \ \ \ \ \ \ \ \ \ \ \ \ \ \ \  \
+\frac{1}{192}\GP{1-\frac{12}{\tau_c}+\frac{54}{\tau_c^2}-\frac{204}
{\tau_c^3}+\frac{477}{\tau_c^4}}\NP{\tau-\tau_c}^4+\cdots\label{E4.4.17}
\end{align}
In order to decide on the number of relevant terms in the expansion, in 
Fig. \ref{F4.4.5} we display $m(\tau)$ calculated by truncating the series at 
different number of terms. 
We see that already up to the fourth order, we obtain a good estimate of the 
exact $m(\tau)$. There is no peak at small times. 
We must mention that $m(\tau)$  (and hence also the modulating function) 
is not symmetric about $\tau=\tau_c$. Hence, if we define the constants 
\begin{align}
m_2&=\frac{1}{8}\GP{1-\frac{3}{\tau_c}}^2,\label{E4.4.18}\\
m_3&=\frac{1}{96}\GP{\frac{36}{\tau_c^2}-\frac{108}{\tau_c^3}},\label{E4.4.19}\\
m_4&=\frac{1}{384}\GP{1-\frac{12}{\tau_c}+\frac{54}{\tau_c^2}-\frac{204}{\tau_c^3}+
\frac{477}{\tau_c^4}},\label{E4.4.20}
\end{align}
then 
\begin{equation}\label{E4.4.21}
m(\tau)=\frac{1/2\sqrt{C}}{1+m_2\NP{\tau-\tau_c}^2+m_3\NP{\tau-\tau_c}^3+m_4\NP{\tau-\tau_c}^4},
\end{equation}
and the modulating function can be written as 
\begin{equation}\label{E4.4.22}
I(\tau)=1+\GP{\frac{D}{2\sqrt{C}}}\,\frac{\cos{\NP{\om_r\tau+\pi/4-\pArg{k_r}}}}{1+m_2\NP{\tau-\tau_c}^2+m_3\NP{\tau-\tau_c}^3+m_4\NP{\tau-\tau_c}^4}.
\end{equation}
In Fig. \ref{F4.4.6}, we compare this expression with that of the modulating function 
given by Eq. \eqref{E4.4.8}. 
In a small region around the critical time, the two expressions coincide exactly 
with small differences away from this region. 
Finally, in Fig. \ref{F4.4.7}, we compare the exact survival probability with that 
using the approximate form of the modulating function. 
Apart from this, the green enveloping curves show the survival probability 
evaluated without the oscillatory part. 
These curves are evaluated by writing the modulating function 
in (\ref{E4.4.14}) with the maximum and minimum values of the cosine term, i.e., 
\begin{equation}\label{E4.4.14p}
I_{\pm}(\tau)=1+D\EP{\frac{e^{-\tau/2}\tau^{-3/2}}{e^{-\tau}+C\tau^{-3}}}\,(\pm 1),
\end{equation}
and $P_{\pm}(\tau)= I_{\pm}(\tau)\MC{P_e(\tau)+P_p(\tau)}$. 
Though somewhat obvious, it is interesting to note that 
the two curves, $P_{+}(\tau)$ and $P_{-}(\tau)$ coincide in all regions except 
for the transition region where they separate. This implies that 
$m(\tau)$ can indeed be used to define the transition region between the 
exponential and the non-exponential region at large times.

\section{Summary and Conclusions}\label{s7}
Writing the survival amplitudes 
based on the Green's function method (GC) as well as the Jost functions 
method (DN), as a Fourier transform similar to the one used in the Fock-Krylov method, 
it is shown that the GC and DN approaches are equivalent up to some constants. 
Such a rewriting allows one to define the densities $\rho^{GC}(E)$ and 
$\rho^{DN}(E)$ which are then compared with the definition of a density 
obtained from a statistical physics motivated expression. The latter is 
obtained from a relation given by Beth and Uhlenbeck which relates the 
density of states, $\rho^{BU}_l(E)$, to the energy derivative of the
scattering phase shift, $d\delta_l/dE$, in the $l^{th}$ partial wave. 
A theorem of Mittag-Leffler further 
allows $\rho^{BU}_l(E)$ to be expressed in terms 
of the poles of the $S$-matrix \cite{ourarxiv}, thus making the comparison with 
$\rho^{GC}(E)$ and $\rho^{DN}(E)$ straightforward. For the case of an isolated 
$s$-wave resonance, $\rho^{GC}(E)$ and $\rho^{DN}(E)$ give the same expression as 
$\rho^{BU}_0(E)$ plus a small correction term. 

Noting that the coefficients appearing in the GC and DN formalism satisfy the same 
conditions, a general analytic form for the survival amplitude in terms of 
the incomplete gamma functions is also derived. 
Apart from this, the analysis for large times is done by 
applying the steepest descent method as well as using the asymptotic 
expansion for the incomplete gamma function. The results 
obtained in both cases are the same, in particular, the $t^{-3}$ power law 
for $s$-wave resonances,  
which is consistent with most of the literature (see however, \cite{rothe,onleykumar} 
for systems without spherical symmetry). 

The equation deduced for the survival amplitude allowed us to easily 
separate the exponential and power law behaviours and define a 
critical time at the intersection of the survival probability for 
intermediate and large times. 
A detailed analysis of the transition region reveals interesting aspects as 
well as the origin of the oscillatory behaviour of the survival probability 
in this region. 
An analytical expression for the critical 
transition time, $\tau_c$, is obtained in terms of the Lambert W function. 
Calculations of $\tau_c$ for the decays which have been measured 
experimentally up to several half-lives with the objective of observing 
the power law behaviour reveal the reason for the negative results of these 
experiments. The number of half-lives after which the power law starts, 
for example, for a narrow nuclear resonance such as $^{56}$Mn is about 
300, whereas the experiment was carried out only up to 45 half-lives. 
However, performing measurements up to 300 half-lives would be practically 
impossible since the exponential decay law would destroy almost all 
the sample by the time the narrow resonance reaches the power law.
Broad resonances such as the $\sigma$ meson reach the power law much 
earlier, however, the lifetime is too short making the experimental 
observation once again difficult. 
The results of the present work indicate that the non-exponential 
behaviour of nuclear and particle resonances at large times is hard to 
observe.  
This conclusion is in agreement with other literature 
such as Ref. \cite{garciaisolated}, where, for the 
case of $^{56}$Mn, the authors found that the 
deviation from the exponential decay law would occur around $\tau_c$ = 331
and in the case of the short lived $^5$He state with a lifetime of 
$\sim$ 10$^{-22}$ s, it would occur around $\tau_c$ = 12.

\begin{acknowledgments}
One of the authors (N. G. K.) acknowledges the support from the 
Faculty of Science, 
Universidad de los Andes, Colombia, through grant no. P18.160322.001-17.
\end{acknowledgments}

\appendix


\section{Evaluation of Survival Amplitude for large times using the 
Steepest Descent Method}\label{A}
The integrals required for the steepest descent method have 
the form\footnote{We follow the notation from Ablowitz and Fokas' book 
for steepest descent method. See \cite{Fokas}, chapter 6.},
\begin{equation}\label{EA1}
\int_C{f(z)e^{t\phi(z)}\,dz},
\end{equation}
where $t>0$ and $t\gg1$, $f(z)$ and $\phi(z)$ are analytic functions in a region 
$D$ and $C\in D$ is the contour (not necessarily closed) of integration. 
In our case, the evaluation of the survival amplitude for large $t$ depends of the computation of the integral $M(k_n,t)$ when $t\gg1$, 
which is identically equal to Eq. \eqref{EA1} if 
$f(z)=(z-k_n)^{-1}$, $\phi(z)=-iz^2$, 
$C=\NL{z\in\C:\pIm{z}=0}$ and $\pIm{k_n}<0$ 
(as of now, we do not include the coefficient $i/2\pi$).

\begin{figure}[htb!]
\centering
\includegraphics{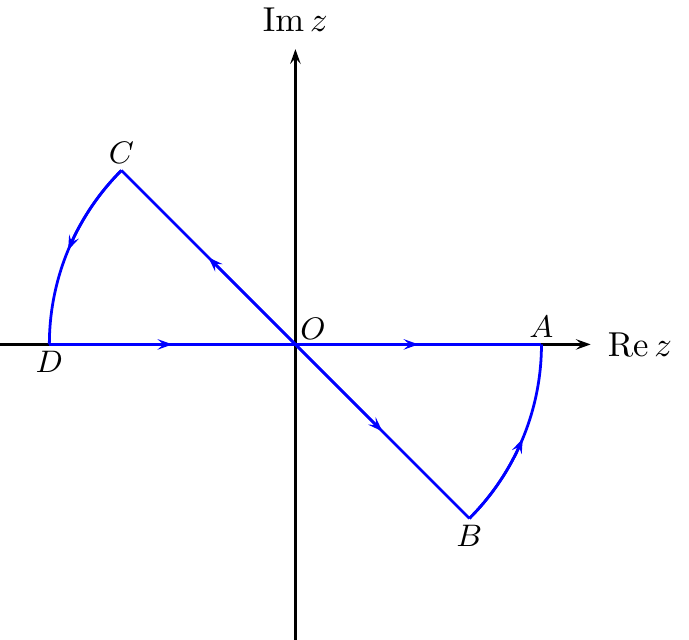}
\caption{Contour of integration for the integral \eqref{E1.3} and the 
directions of steepest descent of its integrand.}\label{FA1}
\end{figure}

The function $\phi(z)$ has one saddle-point of order $n=2$ at $z=0$  
since $\phi'(0)=0$ and $\phi''(0)=-2i=2e^{-i\pi/2}=ae^{\,i\al}$. 
The directions of steepest descent are
\[
\theta=-\frac{\al}{n}+(2m+1)\frac{\pi}{n},\quad m=0,1,\dotsc,n-1.
\]
In our case,
\[
\theta=\frac{3}{4}\pi,\frac{7}{4}\pi.
\]
The contour $C$ must be deformed such that it follows these directions. 
In Fig. \ref{FA1}, we show how the contour $C$ is deformed. 

The line $BOA$ is the contour $C$. 
The straight lines $OC$ and $OB$ are the steepest descent directions. 
For linking the integrals in those contours, we have to close them with 
the arcs of circumference $CD$ and $BA$, both of radius $R$. 
Since the integrand is analytic in the contour $OCD$, 
using the Cauchy theorem, we get, 
\begin{equation}\label{EA2}
\int_{DO}=-\GP{\int_{OC}+\int_{CD}}.
\end{equation}
For the contour $OAB$, however, the integrand has a pole depending on the 
fact if $k_n$ satisfies the condition 
$-\frac{\pi}{4}<\pArg{z}<0$ or not. The residue theorem allows us to write
\begin{equation}\label{EA3}
\int_{OA}=\int_{OB}+\int_{BA}-2\pi ie^{-ik_n^2t}F(k_n),
\end{equation}
where the function $F(k_n)$  is defined by
\begin{equation}\label{EA4}
F(k_n)=
\begin{cases}
1 & -\frac{\pi}{4}<\pArg{z}<0,\\
0 & \text{i. o. c.}
\end{cases}.
\end{equation}
Adding \eqref{EA3} and \eqref{EA4}, we obtain
\begin{equation}\label{EA5}
\int_{DO}+\int_{OA}=\int_{DA}=\int_{OB}+\int_{BA}-\int_{OC}-\int_{CD}-2\pi ie^{-ik_n^2t}F(k_n).
\end{equation}
Since, in the limit $R\to\infty$, both $\int_{BA}$ and $\int_{CD}$ 
tend to zero, \eqref{EA5} takes the following form:
\begin{align}
\intinf{\frac{e^{-itx^2}}{x-k_n}}{x}
&=\int_{\pArg{z}=-\frac{\pi}{4}}{\frac{e^{-itz^2}}{z-k_n}\,dz}-\int_{\pArg{z}=\frac{3\pi}{4}}{\frac{e^{-itz^2}}{z-k_n}\,dz}-2\pi ie^{-ik_n^2t}F(k_n)\notag\\
&=e^{-i\pi/4}\intseminf{\frac{e^{-tr^2}}{re^{-i\pi/4}-k_n}}{r}-e^{\,i3\pi/4}\intseminf{\frac{e^{-tr^2}}{re^{\,i3\pi/4}-k_n}}{r}-2\pi ie^{-ik_n^2t}F(k_n).\label{EA6}
\end{align}
The principal contribution to the value of the integrals for 
$t$ large, comes from a neighbourhood of $r=0$. 
Expanding $\NP{re^{-i\pi/4}-k_n}^{-1}$ and $\NP{re^{\,i3\pi/4}-k_n}^{-1}$ 
in a Taylor series up to the third order around $r=0$ and 
calculating the integrals, we have:
\begin{equation}
\intinf{\frac{e^{-itx^2}}{x-k_n}}{x}=\frac{i\sqrt{\pi}}{k_n}e^{\,i\pi/4}t^{-1/2}-\frac{i\sqrt{\pi}}{2k_n^3}e^{\,i3\pi/4}t^{-3/2}+O(t^{-5/2}),
\end{equation}
and thus,
\begin{equation}\label{EA7}
M(k_n,t)=\frac{i}{2\pi}\intinf{\frac{e^{-itx^2}}{x-k_n}}{x}=-\frac{1}{2\pi}\GC{\frac{1\sqrt{\pi}}{k_n}e^{\,i\pi/4}t^{-1/2}-\frac{\sqrt{\pi}}{2k_n^3}e^{\,i3\pi/4}t^{-3/2}}+O(t^{-5/2}).
\end{equation}
We ignore the exponential term because for large $t$ because it 
is negligible with respect to the negative power of $t$. 
Substituting \eqref{EA7} in \eqref{E1.8} and using properties 
2) and 3) mentioned in section \ref{S1}, we have 
\begin{equation}\label{EA8}
A(t)=\frac{1}{4\sqrt{\pi}}e^{\,i3\pi/4}\GP{\sum_n{\frac{C_n\bar{C}_n}{k_n^3}}}t^{-3/2}+O(t^{-5/2})=
-\frac{1}{\sqrt{4\pi}}e^{\,i\pi/4}\pIm{\GP{\sum_p{\frac{C_n\bar{C}_p}{k_p^3}}}}t^{-3/2}+O(t^{-5/2}).
\end{equation}


\section{Residues of the inverse of the S-matrix}\label{C}
In \cite{ourarxiv}, the authors show that for a system under the influence of 
a central potential of finite range $R$, if the S-matrix for the orbital angular 
momentum $l$ is written as a product, i.e,
\begin{equation}\label{EC34}
S_l(k)=e^{-2iRk}\prod_{n}{\frac{(k+k_{ln})(k-k_{ln}^*)}{(k-k_{ln})(k+k_{ln}^*)}}\prod_{m}{\frac{i\zeta_{lm}+k}{i\zeta_{lm}-k}},
\end{equation}
where $\NL{k_{ln}}$ with 
$n = 1, 2, \dotsc$ corresponding to the resonant poles of the 
S-matrix in the fourth-quadrant 
and $i\zeta_{lm}$, with 
$m =\pm 1,\pm 2, \dotsc$ (with the plus signs corresponding to the 
bound and minus to the virtual states respectively) corresponding to the poles on the 
imaginary axis of the complex $k$ plane, then the residue of $S_l(k)$ at $k=k_{ln}$ is
\begin{equation}\label{EC35}
b_{ln}=\Res{\NC{S_l(k),k=k_{ln}}}=2ik_{ln}e^{-2iRk_{ln}}\tan{\NP{\pArg{k_{ln}}}}\prod_{p\neq n}{\frac{(k_{ln}+k_{lp})(k_{ln}-k_{lp}^*)}{(k_{ln}-k_{lp})(k_{ln}+k_{lp}^*)}}\prod_{m}{\frac{i\zeta_{lm}+k_{ln}}{i\zeta_{lm}-k_{ln}}},
\end{equation} 
and the residue of $S_l(k)$ at $k=-k_{ln}^*$ is
\begin{equation}\label{EC36}
\Res{\NC{S_l(k),k=-k_{ln}^*}}=-b_{ln}^*.
\end{equation}
If the system under consideration has no bound and virtual states, 
Eqs \eqref{EC34} and \eqref{EC35} are simplified to:
\begin{align}\label{EC37q} 
S_l(k)&=e^{-2iRk}\prod_{n}{\frac{(k+k_{ln})(k-k_{ln}^*)}{(k-k_{ln})(k+k_{ln}^*)}},
\\
b_{ln}&=2ik_{ln}e^{-2iRk_{ln}}\tan{\NP{\pArg{k_{ln}}}}\prod_{p\neq n}{\frac{(k_{ln}+
k_{lp})(k_{ln}-k_{lp}^*)}{(k_{ln}-k_{lp})(k_{ln}+k_{lp}^*)}},\label{EC37p}
\end{align}
and Eq. \eqref{EC36} remains the same. 
Here, we are interested in computing the residues of $1/S_l(k)$ at $k=-k_{ln}$ and 
$k=k_{ln}^*$. For the former pole:
\begin{align}
&\Res{\NC{1/S_l(k),k=-k_{ln}}}=\lim_{k\to k_{ln}}{\frac{k+k_{ln}}{S_l(k)}}\notag\\
&=-2k_{ln}\frac{-k_{ln}+k_{ln}^*}{-(k_{ln}+k_{ln}^*)}e^{-2iRk_{ln}}\prod_{p\neq n}{\frac{(-k_{ln}-k_{lp})(-k_{ln}+k_{lp}^*)}{(-k_{ln}+k_{lp})(-k_{ln}-k_{lp}^*)}}\notag\\
&=-2ik_{ln}e^{-2iRk_{ln}}\tan{\NP{\pArg{k_{ln}}}}\prod_{p\neq n}{\frac{(k_{ln}+k_{lp})(k_{ln}-k_{lp}^*)}{(k_{ln}-k_{lp})(k_{ln}+k_{lp}^*)}}=-b_{ln},\label{EC37}
\end{align}
and the latter pole:
\begin{align}
&\Res{\NC{1/S_l(k),k=k_{ln}^*}}=\lim_{k\to k_{ln}}{\frac{k-k_{ln}^*}{S_l(k)}}\notag\\
&=2k_{ln}\frac{-k_{ln}+k_{ln}^*}{k_{ln}+k_{ln}^*}e^{2iRk_{ln}^*}\prod_{p\neq n}{\frac{(k_{ln}^*-k_{lp})(k_{ln}^*+k_{lp}^*)}{(k_{ln}^*+k_{lp})(k_{ln}^*-k_{lp}^*)}}\notag\\
&=\EC{2ik_{ln}e^{-2iRk_{ln}}\tan{\NP{\pArg{k_{ln}}}}\prod_{p\neq n}{\frac{(k_{ln}+k_{lp})(k_{ln}-k_{lp}^*)}{(k_{ln}-k_{lp})(k_{ln}+k_{lp}^*)}}}^*=b_{ln}^*.\label{EC38}
\end{align}
In principle, the residues of the S-matrix depend on all its poles and it is 
difficult to compute them, however, the authors in \cite{ourarxiv} show that it is 
possible to write this residue in two parts such that the former depends only on 
the pole where we calculate the residue and the latter depends on the remaining poles. 
Under certain conditions, the residue can be approximated as follows:
\begin{equation}
b_{ln}\approx2ik_{ln}\tan{\pArg{k_{ln}}}.
\end{equation}
The details of the derivation of this approximation and the conditions 
under which it is valid can be found in \cite{ourarxiv}.

\section{Detailed computation of \texorpdfstring{$\varrho(k)$}{} from \texorpdfstring{$I(k,r,r')$}{}}\label{D}
Substituting \eqref{E19} in \eqref{E9}:
\begin{multline}
\varrho(k)=
\frac{8}{\pi}\,\intdef{0}{R}{\intdef{0}{R}{\psi(r,0)\psi^*(r',0)k^2\,\pRe{\EC{\sum_n{\frac{\iota(k_n,r,r')}{k_n(k^2-k_n^2)}}}}}{r'}}{r}\\
=\frac{4}{\pi}k^2\sum_n{\frac{1}{k_n(k^2-k_n^2)}\intdef{0}{R}{\intdef{0}{R}{\psi(r,0)\psi^*(r',0)\iota(k_n,r,r')}{r'}}{r}}\\+
\frac{4}{\pi}k^2\sum_n{\frac{1}{k_n^*(k^2-{k_n^*}^2)}\intdef{0}{R}{\intdef{0}{R}{\psi(r,0)\psi^*(r',0)\iota^*(k_n,r,r')}{r'}}{r}}.\label{ED20}
\end{multline}
If we write the double integral of the second term of the right side of \eqref{ED20} as the conjugate of some double integral and use the fact that $\iota(k,r,r')=\iota(k,r',r)$, then, 
\begin{align}
\intdef{0}{R}{\intdef{0}{R}{\psi(r,0)\psi^*(r',0)\iota^*(k_n,r,r')}{r'}}{r}
&=\EC{\intdef{0}{R}{\intdef{0}{R}{\psi^*(r,0)\psi(r',0)\iota(k_n,r,r')}{r'}}{r}}^*\notag\\
&=\EC{\intdef{0}{R}{\intdef{0}{R}{\psi^*(r',0)\psi(r,0)\iota(k_n,r',r)}{r'}}{r}}^*\notag\\
&=\EC{\intdef{0}{R}{\intdef{0}{R}{\psi(r,0)\psi^*(r',0)\iota(k_n,r,r')}{r'}}{r}}^*,\label{ED21}
\end{align}
and substituting \eqref{ED21} in \eqref{ED20}, we get:
\begin{equation}\label{ED22}
\varrho(k)=
k^2\,\pRe{\sum_n{\frac{1}{k_n(k^2-k_n^2)}\GC{\frac{8}{\pi}\intdef{0}{R}{\intdef{0}{R}{\psi(r,0)\psi^*(r',0)\iota(k_n,r,r')}{r'}}{r}}}}.
\end{equation}
Taking the integral in the square brackets and defining 
\begin{equation}\label{ED23}
a(k_n)\equiv{4}{i}\intdef{0}{R}{\intdef{0}{R}{\psi(r,0)\psi^*(r',0)\iota(k_n,r,r')}{r'}}{r},
\end{equation}
allows us to write the Eq. \eqref{ED22} in the following form:
\begin{equation}\label{ED24}
\varrho(k)=
\frac{2}{\pi}\,k^2\,\pRe{\sum_n{\frac{ia(k_n)}{k_n(k_n^2-k^2)}}}.
\end{equation}


\section{Evaluation of integrals for computing the survival amplitude}\label{B}
\begin{figure}[htb!]
\centering
\includegraphics{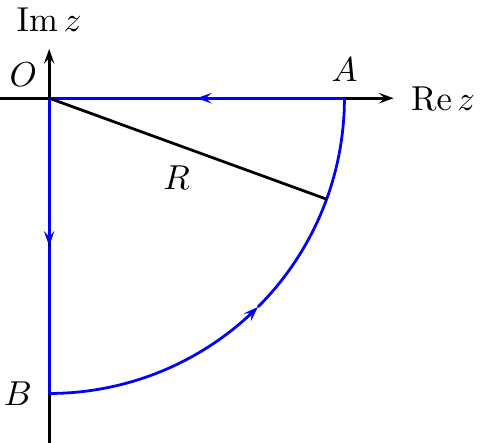}
\caption{Contour of integration for computing the integral \eqref{EB1}.}\label{FB1}
\end{figure}

For calculating the survival amplitude given by \eqref{E3.1}, 
we need to study the integral
\begin{equation}\label{EB1}
I(\al)=\oint_C{\frac{\sqrt{z}}{z-\al}e^{-itz}\,dz},
\end{equation}
where $C$ is the contour shown in Fig. \ref{FB1} and $t>0$.

Since the integrand has a branch point at $z=0$, we take the principal 
branch of $\sqrt{z}$ such that $|\pArg{z}|<\pi$. 
If $\al$ satisfies $\pArg{\al}\in(-\frac{\pi}{2},0)$ and $R<|\al|$, using 
the residue theorem, we have: 
\begin{equation}\label{EB2}
I(\al)=2\pi i\al^{1/2}e^{-i\al t}.
\end{equation}
If $\al$ is not inside or on the contour, the Cauchy's theorem leads us to,
\begin{equation}\label{EB3}
I(\al)=0.
\end{equation}
Both expressions can be written together as
\begin{equation}\label{EB4}
I(\al)=
\begin{cases}
2\pi i\al^{1/2}e^{-i\al t}, & \pRe{\al}\neq0,\pIm{\al}\neq0,\pArg{\al}\in(-\frac{\pi}{2},0),\\
0 & \pRe{\al}\neq0,\pIm{\al}\neq0,\pArg{\al}\notin(-\frac{\pi}{2},0).
\end{cases}
\end{equation}
On the other hand,
\[
I(\al)=\int_{AO}+\int_{OB}+\int_{BA}.
\]
For the segment $AO$, $z=x$. When $R\to\infty$,
\[
\int_{AO}=-\int_{OA}=-\int_{0}^{R}{\frac{\sqrt{x}}{x-\al}\,e^{-itx}\,dx}\to-\intseminf{\frac{\sqrt{x}}{x-\al}\,e^{-itx}}{x}.
\]
For the segment $OB$, $z=-iy$. When $R\to\infty$,
\[
\int_{BO}=\int_{0}^{R}{\frac{\sqrt{-iy}}{-iy-\al}e^{-ty}\,(-idy)}\to
e^{-i\pi/4}\intseminf{\frac{\sqrt{y}}{y-i\al}\,e^{-ty}}{y}.
\]
For the segment $BA$, let $z=Re^{-i\theta}$, where $\theta\in\NC{0,\frac{\pi}{2}}$. The integral on this segment is:
\[
\int_{BA}=-\int_{0}^{\pi/2}{\frac{R^{1/2}e^{-i\theta/2}}{Re^{-i\theta}-\al}{\,}e^{-tR\sin{\theta}-itR\cos{\theta}}\NP{-iRe^{-i\theta}}\,d\theta}.
\]%
Taking the modulus of this integral and supposing $R>|\al|$, we have:
\[
\GB{\int_{BA}}\leq\intdef{0}{\pi/2}{\frac{R^{3/2}}{R-|\al|}\,e^{-tR\sin{\theta}}}{\theta}
\leq\frac{R^{3/2}}{R-|\al|}\intdef{0}{\pi/2}{\,e^{-2Rt{\theta}\pi}}{\theta}=
\frac{\pi}{2t}\frac{R^{1/2}}{R-|\al|}\NP{1-e^{-Rt}}.
\]
Here, we used $\sin{\theta}\geq2\theta/\pi$ with $\theta\in\NC{0,\frac{\pi}{2}}$. When $R\to\infty$, 
\[
\GB{\int_{BA}}\to0,
\]
and,
\[
{\int_{BA}}\to0.
\]
In the limit $R\to\infty$, $I(\al)$ is equal to
\[
I(\al)=-\intseminf{\frac{\sqrt{x}}{x-\al}\,e^{-itx}}{x}+e^{-i\pi/4}\intseminf{\frac{\sqrt{y}}{y-i\al}\,e^{-ty}}{y}.
\]
The integral \cite{Bateman1}
\[
\intseminf{\frac{E^{\nu}}{E+\sig}e^{-sE}}{E}=\Ga(\nu+1)e^{\sig s}\sig^{\nu}\Ga(-\nu,\sig s),\quad\pRe{\nu}>-1,\pRe{s}>0,|\pArg{\sigma}|<\pi,
\]
where $\Ga(\al,z)$ is the incomplete gamma function \cite{Lebedev},
allows us to write $I(\al)$ as
\begin{equation}\label{EB5}
I(\al)=-\intseminf{\frac{\sqrt{x}}{x-\al}\,e^{-itx}}{x}+e^{-i\pi/4}\Ga\NP{\tfrac{3}{2}}e^{-i\al t}\NP{-i\al}^{1/2}\Ga\NP{-\tfrac{1}{2},-i\al t}.
\end{equation}
Finally, 
\begin{equation}\label{EB6}
\intseminf{\frac{\sqrt{x}}{\al-x}\,e^{-itx}}{x}=I(\al)+i\frac{\sqrt{\pi}}{2}\,\al^{1/2}e^{-i\al t}\Ga\NP{-\tfrac{1}{2},-i\al t},\quad|\pArg{\al}|<\pi.
\end{equation}

\begin{thebibliography}{99}
\bibitem{khalfin}
L. A. Khalfin, Zh. Eksp. Teor. Fiz. {\bf 33}, 1371 (1957). 
\bibitem{urbanowski}
K. Urbanowski, Acta Phys. Pol. B {\bf 48} 1847 (2017); {\it ibid}, Eur. Phys. J D
{\bf 71}, 118 (2017).
\bibitem{fonda}
L. Fonda, G. C. Ghirardi and A. Rimini, Rep. Prog. Phys. {\bf 41}, 587 (1987).
\bibitem{giraldi}
F. Giraldi, Eur. Phys. J. D {\bf 70} 229 (2016).  
\bibitem{shorttime}
J. Levitan, Phys. Lett. A {\bf 129}, 267 (1988); 
H. Nakazato and S. Pascazio, Mod. Phys. Lett. A {\bf 10}, 3103 (1995).
\bibitem{norman}
E. B. Norman, S. B. Gazes, S. G. Crane and D. A. Bennett, Phys. Rev. Lett. 
{\bf 60}, 2246 (1988). 
\bibitem{rothe}
C. Rothe, S. I. Hintschich and A. P. Monkman, Phys. Rev. Lett. {\bf 96}, 163601 
(2006).
\bibitem{lawrence}
J. Lawrence, J. Opt. B: Quantum Semiclass. Opt. {\bf 4}, S446 (2002). 
\bibitem{Fock1}
V. Fock and N. Krylov, JETP {\bf 17}, 93 (1947).
\bibitem{Sym1}
G. Garc\'ia-Calder\'on, {\it Resonant States and the Decay Process}, 
``Symmetries in Physics", eds. A. Frank and K. B. Wolf, 
Springer-Verlag, p. 252-272 (1992).  
\bibitem{Garcia2}
G. Garc\'ia-Calder\'on, J. L. Mateos and M. Moshinsky, 
Phys. Rev. Lett. {\bf 74}, 337 (1995).
\bibitem{Nakazato}
H. Nakazato, M. Namiki and S. Pascazio, Int. J. Mod. Phys. B {\bf 10}, 247 (1996). 
\bibitem{dijknogaPRL}
W. van Dijk and Y. Nogami, Phys. Rev. Lett. {\bf 83}, 2867 (1999). 
\bibitem{dijknogaPRC}
W. van Dijk and Y. Nogami, Phys. Rev. C {\bf 65}, 024608 (2002). 
\bibitem{Garcia1}
G. Garc\'ia-Calder\'on, J. L. Mateos and M. Moshinsky, Annals of Phys. {\bf 249}, 430 
(1996).
\bibitem{AdvChemGarcia}
G. Garc\'ia-Calderon, Advances in Quantum Chemistry {\bf 60}, 407 (2010). 
\bibitem{AIPGarcia}
G. Garc\'ia-Calderon, AIP Conf. Proc. {\bf 1334}, 84 (2011). 
\bibitem{garciaisolated}
G. Garc\'ia-Calderon, V. Riquer and R. Romo, J. Phys. A {\bf 34}, 4155 (2001). 
\bibitem{GarciaPRA76}
G. Garc\'ia-Calderon, I. Maldonado and J. Villavicencio, Phys. Rev. A {\bf 76}, 012103 
(2007).
\bibitem{GarciaPRA88}
G. Garc\'ia-Calder\'on, I. Maldonado and J. Villavicencio, Phys. Rev. {\bf 88}, 
052114 (2013).
\bibitem{dijkPRE93}
Wytse van Dijk, Phys. Rev. E {\bf 93}, 063307 (2016).
\bibitem{CavalPRL80}
R. M. Cavalcanti, Phys. Rev. Lett. {\bf 80}, 4353 (1998).
\bibitem{garciaPRL80}
G. Garc\'ia-Calder\'on, J. L. Mateos and M. Moshinsky, Phys. Rev. Lett. {\bf 80}, 
4354 (1998).
\bibitem{dijkPRL901}
W. van Dijk and Y. Nogami, Phys. Rev. Lett. {\bf 90}, 028901 (2003). 
\bibitem{garciaPRL90}
G. Garc\'ia-Calder\'on, J. L. Mateos and M. Moshinsky, Phys. Rev. Lett. {\bf 90}, 
028902 (2003). 
\bibitem{Kelkar1}
N. G. Kelkar and M. Nowakowski, J. Phys. A {\bf 43}, 385308 (2010).
\bibitem{Kelkar2}
N. G. Kelkar, M. Nowakowski and K. P. Khemchandani, Phys. Rev. C {\bf 70}, 024601 
(2004).
\bibitem{ourarxiv}
D. F. Ram\'irez Jim\'enez and N. G. Kelkar, Ann. Phys. {\bf 396}, 18 (2018); 
arXiv:1802.09467 (2018). 
\bibitem{wigner55}
E. P. Wigner, Phys. Rev. {\bf 98}, 145 (1955).
\bibitem{smith}
F. T. Smith, Phys. Rev. {\bf 118}, 349 (1960).
\bibitem{meandMPRA}
N. G. Kelkar and M. Nowakowski, Phys. Rev. A {\bf 78}, 012709 (2008). 
\bibitem{mePRL}
N. G. Kelkar, Phys. Rev. Lett. {\bf 99}, 210403 (2007).
\bibitem{bethuhl}
E. Beth and G. E. Uhlenbeck, Physica {\bf 4}, 915 (1937).
\bibitem{huang}
K. Huang, {\it Statistical Mechanics}, Wiley, New York (1987).
\bibitem{dashen1}
R. F. Dashen, S. Ma and H. J. Bernstein, Phys. Rev. {\bf 137}, 345 (1969).
\bibitem{dashen2}
R. F. Dashen and R. Rajaraman, Phys. Rev. D {\bf 10}, 708 (1974).
\bibitem{Sym2}
G. Garc\'ia-Calder\'on, G. Loyola and M. Moshinsky, 
{\it The Decay Process: An Exactly Soluble Example and its Implications}, 
``Symmetries in Physics", eds. A. Frank and K. B. Wolf, 
Springer-Verlag, p. 273-292 (1992).  
\bibitem{kangoller}
Xian-Wei Kang and J. A. Oller, Eur. Phys. J. C {\bf 77}, 399 (2017). 
\bibitem{Garcia3}
G. Garc\'ia-Calder\'on and R. Peierls, Nucl. Phys. A {\bf 265}, 443 (1976).
\bibitem{bogda}
J. Bogdanowicz, M. Pindor and R. Raczka, Found. Phys. {\bf 25}, 833 (1995).
\bibitem{zeldovich}
Y. B. Zeldovich, JETP, {\bf 12}, 542 (1961).
\bibitem{Gareev}
J . Bang, F. A. Gareev, M. H. Gizzatkulov and S. A. Gonchanov, 
Nucl. Phys. A {\bf 309}, 381 (1978). 
\bibitem{onleykumar}
D. S. Onley and A. Kumar, Am. J. Phys. {\bf 60}, 432 (1992). 
\bibitem{moshPRA8488}
M. Moshinsky, Phys. Rev. {\bf 84}, 525 (1951); {\it ibid} {\bf 88}, 625 (1952); 
G. Garc\'ia-Calderón and A. Rubio, Phys. Rev. A {\bf 55}, 3361 (1997).
\bibitem{joachain}
C. J. Joachain,
{\it Quantum Collision Theory} (North-Holland, Amsterdam 1975). 
\bibitem{cavalcanti}
R. M. Cavalcanti and C. A. A. de Carvalho, Revista Brasileira de Ensino de F\'isica 
{\bf 21}, 464 (1999).
\bibitem{GarciaPhysScripta}
G. Garc\'ia-Calder\'on, A. M\'attar and J. Villavicencio, Phys. Scr. T {\bf 151}, 014076 
(2012).  
\bibitem{Zeldovich2}
A.I. Baz, Ya. B. Zeldovich, A.M. Perelomov, {\it Scattering, Reactions and Decay in Nonrelativistic Quantum Mechanics}, Israel
Program for Scientific Translations, Springfield, 1969.
\bibitem{Sitenko} A. G. Sitenko, {\it Scattering Theory}, Springer-Verlag, 1991.
\bibitem{rakitelander}
S. A. Rakityansky and N. Elander, J. Phys. A {\bf 45}, 135209 (2012). 
\bibitem{MittagL}
E. T. Copson, 
{\it An Introduction to the Theory of Functions of a Complex Variable}, 
Oxford University Press (1935).
\bibitem{Tijonov} A. G. Sveshnikov and A. N. Tikhonov, {\it The Theory of Functions of a Complex Variable}, Mir Publishers, 1974.
\bibitem{polaco}
A. Brzeski and J. Lukierski, Acta Physica Polonia, Vol {\bf B6}, 577 (1975). 
\bibitem{urbanowski2018}
K. Raczynska and K. Urbanowski, preprint, arXiv:1802.01441 (2018).
\bibitem{Lebedev}
N. N. Lebedev, {\it Special functions and their applications}, Dover Publications Inc. 
(1975).
\bibitem{Copson2}
E. T. Copson, {\it Asymptotic Expansions}, Cambridge University Press (1965). 
\bibitem{winter1962}
R. G. Winter, Phys. Rev. {\bf 126}, 1152 (1962).
\bibitem{pdg}
C. Patrignani et al., Chin. Phys. C {\bf 40}, 100001 (2016).
\bibitem{Garcia2006}
G. Garc\'ia-Calderon and J. Villavicencio, Phys. Rev. A {\bf 73}, 062115 (2006).
\bibitem{Garcia2016}
G. Garc\'ia-Calderon and R. Romo, Phys. Rev. A {\bf 93}, 022118 (2016). 
\bibitem{rutherford}
E. Rutherford, Stizungsber. Akad. Wiss. Wien, Math.-Naturwiss. Kl., Abt. 2A {\bf 120}, 
303 (1911).
\bibitem{buttwilson}
D. K. Butt and A. R. Wilson, J. Phys. A {\bf 5}, 1248 (1972). 
\bibitem{nikolaev}
N. N. Nikolaev, Usp. Fiz. Nauk {\bf 95}, 506 (1968) [Sov. Phys. Usp. {\bf 11}, 522 
(1968)]. 
\bibitem{winter}
R. G. Winter, Phys. Rev. {\bf 123}, 1503  (1961).
\bibitem{Fokas}
M. J. Ablowitz and A. S. Fokas, {\it Complex Variables: Introduction and Applications}, 
Cambridge University Press, 2$^{nd}$ edition (2003).
\bibitem{Bateman1}
A. Erd\'erly, {\it Table of Integral Transforms}, Vol. I, McGrawHill (1954). 

\end{thebibliography}

\end{document}